\documentclass[a4paper,12pt]{article}
\usepackage{theorem}

\newtheorem{theorem}{Theorem}[section]

\newtheorem{lemma}[theorem]{Lemma}

\newcount\driver
\newcount\bozza

\font\msytw=msbm10 scaled\magstep1
\font\msytww=msbm8 scaled\magstep1

\font\indbf=cmbx10 scaled\magstep2

{\count255=\time\divide\count255 by 60
\xdef\hourmin{\number\count255}
        \multiply\count255 by-60\advance\count255 by\time
   \xdef\hourmin{\hourmin:\ifnum\count255<10 0\fi\the\count255}}

\let\a=\alpha \let\b=\beta    \let\g=\gamma     \let\d=\delta     \let\e=\varepsilon
  \let\h=\eta     \let\th=\vartheta \let\k=\kappa     \let\l=\lambda
\let\m=\mu    \let\n=\nu      \let\x=\xi        \let\p=\pi        \let\r=\rho
\let\s=\sigma \let\t=\tau            \let\c=\chi
\let\ps=\psi   \let\o=\omega     
 \let\D=\Delta       \let\L=\Lambda    
             
\let\O=\Omega 

\def\VV{{\cal V}}
\def\WW{{\cal W}}

\def\BB{{\cal B}}
\def\LL{{\cal L}}
\def\DD{{\cal D}}

\def\pp{{\bf p}}\def\qq{{\bf q}}\def\xx{{\bf x}}
\def\yy{{\bf y}}\def\kk{{\bf k}}\def\nn{{\bf n}}
\def\zz{{\bf z}}

       \def\oo{{\underline \omega}}
\def\ee{{\underline \varepsilon}}  
          
          \def\ux{{\underline\xx}}
\def\uk{{\underline \kk}}

\def\u0{{\underline 0}}

\def\RRR{\hbox{\msytw R}}

        \def\ZZZ{\hbox{\msytw Z}}
\def\zzzz{\hbox{\msytww Z}}

\let\dpr=\partial

\let\==\equiv

\let\io=\infty
\let\0=\noindent

\def\*{\vspace{.3cm}}

\def\eg{\hbox{\it e.g.\ }}

\def\tilde#1{{\widetilde #1}}

\def\lft{\left}
\def\rgt{\right}

\def\la{{\langle}}
\def\ra{{\rangle}}

\def\tende#1{\,\vtop{\ialign{##\crcr\rightarrowfill\crcr
             \noalign{\kern-1pt\nointerlineskip}
             \hskip3.pt${\scriptstyle #1}$\hskip3.pt\crcr}}\,}
\def\otto{\,{\kern-1.truept\leftarrow\kern-5.truept\to\kern-1.truept}\,}
\def\fra#1#2{{#1\over#2}}

\def\wh#1{\widehat{#1}}
\def\hat#1{\wh{#1}}
\def\sqt[#1]#2{\root #1\of {#2}}

\def\ha{{\widehat \a}}

\def\hp{{\widehat \ps}}

\def\VV{{\cal V}}
\def\WW{{\cal W}}
\def\BB{{\cal B}}
\def\LL{{\cal L}}
\def\DD{{\cal D}}

\def\T#1{{#1_{\kern-3pt\lower7pt\hbox{$\widetilde{}$}}\kern3pt}}
\def\VVV#1{{\underline #1}_{\kern-3pt
\lower7pt\hbox{$\widetilde{}$}}\kern3pt\,}
\def\W#1{#1_{\kern-3pt\lower7.5pt\hbox{$\widetilde{}$}}\kern2pt\,}

\def\indica{\leaders \hbox to 0.5cm{\hss.\hss}\hfill}
\def\guida{\leaders\hbox to 1em{\hss.\hss}\hfill}
\mathchardef\oo= "0521

\def\pp{{\bf p}}\def\qq{{\bf q}}\def\xx{{\bf x}}
\def\yy{{\bf y}}\def\kk{{\bf k}}\def\nn{{\bf n}}
\def\zz{{\bf z}}

\def\oo{{\underline \omega}}

\def\qed{\raise1pt\hbox{\vrule height5pt width5pt depth0pt}}

\def\indic{\hbox{\raise-2pt \hbox{\indbf 1}}}

\def\RRR{\hbox{\msytw R}}

 \def\ZZZ{\hbox{\msytw Z}}
\def\zzzz{\hbox{\msytww Z}}

\def\ins#1#2#3{\vbox to0pt{\kern-#2 \hbox{\kern#1 #3}\vss}\nointerlineskip}

\newdimen\xshift \newdimen\xwidth \newdimen\yshift
\newcount\griglia

\def\insertplot#1#2#3#4#5#6{%
\xwidth=#1pt \xshift=\hsize \advance\xshift by-\xwidth \divide\xshift by 2%
\begin{figure}[ht]
\vspace{#2pt} \hspace{\xshift}
\begin{minipage}{#1pt}
#3 \ifnum\driver=1 \griglia=#6
\ifnum\griglia=1 \openout13=griglia.ps \write13{gsave .2
setlinewidth} \write13{0 10 #1 {dup 0 moveto #2 lineto } for}
\write13{0 10 #2 {dup 0 exch moveto #1 exch lineto } for}
\write13{stroke} \write13{.5 setlinewidth} \write13{0 50 #1 {dup 0
moveto #2 lineto } for} \write13{0 50 #2 {dup 0 exch moveto #1
exch lineto } for} \write13{stroke grestore} \closeout13
\includegraphics{griglia.ps} \fi
\includegraphics{#4.ps}\fi%
\ifnum\driver=2 \fi
\end{minipage}
\caption{#5}
\end{figure}
}

\newdimen\shift \shift=-1.5truecm
\def\lb#1{%
\ifnum\bozza=1
\label{#1}\rlap{\hbox{\hskip\shift$\scriptstyle#1$}}
\else\label{#1} \fi}

\def\be{\begin{equation}}
\def\ee{\end{equation}}
\def\bea{\begin{eqnarray}}\def\eea{\end{eqnarray}}
\def\bean{\begin{eqnarray*}}\def\eean{\end{eqnarray*}}
\def\bfr{\begin{flushright}}\def\efr{\end{flushright}}
\def\bc{\begin{center}}\def\ec{\end{center}}
\def\bal{\begin{align}}\def\eal{\end{align}}
\def\ba#1{\begin{array}{#1}} \def\ea{\end{array}}
\def\bd{\begin{description}}\def\ed{\end{description}}
\def\bv{\begin{verbatim}}\def\ev{\end{verbatim}}
\def\nn{\nonumber}
\def\Halmos{\hfill\vrule height10pt width4pt depth2pt \par\hbox to \hsize{}}
\def\pref#1{(\ref{#1})}
 
\def\virg{\quad,\quad}

\begin{document}
\title{Universality relations in non-solvable quantum spin chains}

\author{
G. Benfatto\thanks{Dipartimento di Matematica, Universit\`a di Roma ``Tor
Vergata'', Via della Ricerca Scientifica, I-00133, Roma}
\\e-mail: benfatto@mat.uniroma2.it,
\and V. Mastropietro{${}^\ast$}
\\e-mail: mastropi@mat.uniroma2.it
}

\date{11 July 2009}
\maketitle

\begin{abstract}
We prove the exact relations between the critical exponents and the
susceptibility, implied by the Haldane Luttinger liquid conjecture, for a
generic lattice fermionic model or a quantum spin chain with short range weak
interaction. The validity of such relations was only checked in some special
solvable models, but there was up to now no proof of their validity in
non-solvable models.
\end{abstract}

\section{Introduction and Main results}

One dimensional (1D) electron systems can be experimentally realized
\cite{[1],[2]} and their properties can be measured with increasing precision.
Realistic models are very difficult to study and most of the theoretical
predictions for such systems (for some recent experiments see \cite{[3]}) are
based on a number of {\it conjectures}, whose mathematical proof is quite hard.

Kadanoff \cite{[4]} and  Luther and Peschel \cite{[5]} proposed that a large
class of interacting 1D fermionic systems, quantum spin chains or 2D spin
systems belongs to the same universality class. The critical indices appearing
in the correlations are not the same (on the contrary, the indices depend on
all details of the Hamiltonian), but they verify universal {\it extended
scaling relations} between them,  with the effect that all indices can be
expressed in terms of any one of them. Usually, such hypothesis is formulated
by saying that there exists a quantity $K$, whose value depends on the model,
such that the critical indices can be expressed by simple universal relations
in terms of $K$. The validity of such relations can be verified in the {\it
Luttinger model}, which was solved by Mattis and Lieb \cite{[7]}.

Haldane \cite{[5]} observed that in general even the knowledge of a single
exponent is lacking, while the thermodynamic quantities are usually much more
accessible, both experimentally and theoretically. He conjectured that certain
relations between the parameter $K$ and thermodynamical quantities, like the
compressibility, are universal properties in a large class of models which he
named {\it Luttinger liquids}. In the case of models which can be analyzed by
{\it Bethe ansatz} and belonging to such class, the Haldane conjecture allows
the exact computation of critical indices; indeed the Bethe ansatz by itself
allows only the (partially rigorous) computation of spectral properties but not
of the exponents.

The Haldane relations can be verified in the case of the {\it Luttinger model},
where the exact solution of \cite{[7]} allows to calculate all the spectral
quantities and the correlations. In the case of the XYZ spin chain model, whose
ground state energy can be computed by the Bethe ansatz \cite{[9]}, the
relations can be verified {\it assuming} the validity of the Kadanoff extended
relations. The Haldane conjecture, stating that such relations should be valid
in a general class of models (solvable or non-solvable) has been the subject of
an impressive number of studies, see \eg \cite{[7a]} for a review; we mention
the RG analysis in \cite{[8]} (valid only for the Luttinger model) and the
(heuristic) probabilistic approach in \cite{[11]}. While such analyses give
deep insights, a proof of the conjecture for generic non-solvable models is
still lacking.

In recent times, some of the Kadanoff relations have been proved in \cite{[14]}
for several (solvable and non-solvable) planar spin models, by rigorous
Renormalization Group methods. In this paper we will extend such results to
prove one of the Haldane relations for generic non-solvable lattice fermionic
models or quantum spin chains with short range weak interaction. For
definiteness (but our results, as it is evident from the proof, could be easily
extended to 1-d fermionic continuum models) we consider a quantum spin chain
with a non local interaction, whose Hamiltonian is
\be\lb{1.1} H=-\sum_{x=1}^{L-1} [J_1 S^1_x S^1_{x+1}+J_2 S^2_x
S^2_{x+1}] -h \sum_{x=1}^L S^3_x +\l \sum_{1\le x,y\le L} v(x-y)
S^3_x S^3_{y}+ U^1_L\;, \ee
where $S^\a_x = \s^\a_x/2$ for $i=1,2,\ldots,L$ and $\a=1,2,3$,
$\s^\a_x$ being the Pauli matrices, and $U^1_L$, to be fixed
later, depends on the boundary conditions; finally $v(x-y)=v(y-x)$
and $|v(x-y)|\le C e^{-\k |x-y|}$. If $v(x-y)=\d_{|x-y|,1}/2$ and
$h=0$, \pref{1.1} is the hamiltonian of the $XXZ$ spin chain in a
zero magnetic field, which can be diagonalized by the Bethe ansatz
\cite{[9]}; the same is true for the general $XYZ$ model, always
for $h=0$ \cite{[15]}, but  in the other cases no exact solution
is known.

%

It is well known that the operators $a_x^\pm \= \prod_{y=1}^{x-1}
(-\s_y^3) \s_x^\pm$ are a set of anticommuting operators and that,
if $\s_x^{\pm}=(\s_x^1\pm i \s_x^2)/2$, we can write
\be \s^-_x=e^{-i\pi \sum_{y=1}^{x-1} a^+_y a^-_y }  a^-_x\;, \quad
\s^+_x= a^+_x e^{i\pi\sum_{y=1}^{x-1} a^+_y a^-_y }\;,\quad
\s^3_x=2 a^+_x a^-_x-1\;. \ee
Hence, if we fix the units so that $J_1=J_2=1$ we get
\bea\lb{z} && H = -\sum_{x=1}^{L-1} \fra12 [ a^+_{x} a^-_{x+1}+
a_{x+1}^+
a^-_{x}] -h\sum_{x=1}^L ( a^+_x a^-_x-{1\over 2}) +\nn\\
&& +\l \sum_{1\le x,y\le L} v(x-y)( a^+_x a^-_x-{1\over 2})(
a^+_{y} a^-_{y}-{1\over 2}) +U_L^2\;, \eea
where $U_L^2$ is the boundary term in the new variables. We choose
it so that the fermionic Hamiltonian coincides with the
Hamiltonian of a fermion system on the lattice with periodic
boundary conditions.

If $O_x$ is a local monomial in the $S^\a_x$ or $a^\pm_x$ operators, we call
$O_\xx=e^{H x_0} O_x e^{-H x_0}$ where $\xx=(x,x_0)$; moreover, if
$A=O_{\xx_1}\cdots O_{\xx_n}$,  $<A>_{L,\b}= Tr[e^{-\b H}{\bf T}(A)]/ Tr[e^{-\b
H}]$, ${\bf T}$ being the time order product, denotes its expectation in the
grand canonical ensemble, while $<A>_{T;L,\b}$ denotes the corresponding
truncated expectation. We will use also the notation $<A>_{T}=\lim_{L,\b\to\io}
<A>_{T;L,\b}$.

In recent times, constructive Renormalization Group techniques, combined with
asymptotic Ward Identities, have been applied to the $XYZ$ model
\cite{[13],[13a]}. The extension to the general spin chain model \pref{1.1} is
immediate and one can prove that, for small $\l$, $J_1=J_2=1$ and large $\xx$,
\be\lb{8a} \la a^-_\xx a^+_{\bf 0}\ra_T\sim
g_0(\xx){1+\l f(\l)\over (x_0^2+v_s^2 x^2)^{(\eta/2)}}\;, \ee
where $f(\l)$ is a bounded function, $\eta=a_0\l^2+O(\l^3)$, with
$a_0>0$, and
\be\lb{8c} g_0(\xx)= \sum_{\o=\pm}{e^{i\o p_F x}\over -i x_0+ \o
v_s x}\;, \ee
\be v_s=v_F+O(\l)\virg p_F=\cos^{-1}(h+\l)+O(\l) \virg v_F=\sin
p_F\;. \ee
From \pref{8a} we see that the interaction has two main effects. The first one
is to change the value of the Fermi momentum from $\cos^{-1} (h)$ to $p_F$ and
the sound velocity from $v_F$ in the non interacting case to $v_s$. The second
effect is that the power law decay is changed; the 2-point function is
asymptotically given by the product of the non-interacting one (with a
different sound velocity) times an extra power law decay factor with
non-universal index $\eta$.

It was also proved in \cite{[13],[13a]} that the spin-spin correlation in the
direction of the 3-axis (or, equivalently, the fermionic density-density
correlation) is given, for large $\xx$, by
\be\lb{a1} \la S^{(3)}_\xx S^{(3)}_{\bf 0}\ra_T\sim \cos(2p_F x)
\O^{3,a}(\xx) +\O^{3,b}(\xx)\;, \ee
\bea
\O^{3,a}(\xx) &=& {1+A_1(\xx)\over 2\p^2[x^2+(v_s x_0)^2]^{X_+}}\;,\\
\O^{3,b}(\xx) &=& {1\over 2\p^2[x^2+(v_s x_0)^2]}\Big\{
{x_0^2-(x/v_s)^2 \over x^2+(v_s x_0)^2} + A_2(\xx)\Big\}\;, \eea
with $|A_1(\xx)|, |A_2(\xx)|\le C|\l|$ and $X_+=1-a_1 \l+O(\l^2)$, $a_1>0$.
Finally, by using the results of \cite{[16]}, one can prove that the Cooper
pair density correlation, that is the correlation of the operator
$\r^{c}_\xx=a^+_\xx a^+_{\xx'}+a^-_\xx a^-_{\xx'}$, $\xx'=(x+1,x_0)$, behaves
as
\begin{equation}\lb{a2}
\la \r^{c}_\xx \r^{c}_{\bf 0}\ra_T\sim {1+A_3(\xx)\over
2\pi^2(x^2+v_s^2 x_0^2)^{X_-}}\;,
\end{equation}
with $X_-=1+a_1 \l+O(\l^2)$, $a_1$ being the same constant
appearing in the first order of $X_+$.

In the case $J_1\not= J_2$ the correlations decay faster than any power with
rate $\x$ such that
\be \x\sim C |J_1-J_2|^{\bar\n}\;, \ee
with $\bar\n= 1+a_1 \l+O(\l^2)$, $a_1$ being again the same constant appearing
in the first order of $X_+$.

Several physical quantities are expressed in terms of the Fourier transform of
the correlations; in particular, if we call
\be\lb{FTO}
\hat\O(\pp)=\lim_{\b,L\to\io}\int_{-\b/2}^{\b/2}
dx_0\sum_{x\in\L} e^{i\pp\xx} \la S^{(3)}_\xx S^{(3)}_{\bf 0}\ra_{T;L,\b}\;,
\ee
the {\it susceptibility} is given by
\be\lb{kk} \k=\lim_{p\to 0}\hat\O(0,p)\;.
\ee
Note that, in the fermion system, $\k= \k_c \r^2$, where $\k_c$ is the {\it
fermionic compressibility} and $\r$ is the fermionic density, see \eg (2.83) of
\cite{[8a]} or (3.16) of \cite{[8]}.

Our results can be summarized by the following theorem.
\begin{theorem}\lb{thm1} For small $\l$ there exists an analytic function
$K(\l)$ such that
\bea\lb{xxx} &&X_+=K\virg X_-=K^{-1}\;,\\
\lb{xxxx}&&\bar\n={1\over 2-K^{-1}}\virg 2\eta=K+K^{-1}-2\;, \eea
with
\be\lb{l} K=1- \l {\hat v(0)- \hat v(2 p_F) \over \pi \sin
p_F}+O(\l^2)\;. \ee
Moreover,
\be\lb{fon5} \hat\O(\pp)= {K\over \pi v_s}{v_s^2p^2\over
p_0^2+v_s^2 p^2}+R(\pp)\;,
\ee
%
with $R(\pp)$ continuous and such that $R({\bf 0})=0$, so that
\be\lb{gio} \k={1\over\pi}{K\over v_s}\;. \ee
\end{theorem}

The relations \pref{xxx} are the extended scaling laws conjectured
by Kadanoff \cite{[4]} and Luther and Peschel \cite{[5]}. The
critical indices, as functions of $\l$, are non-universal and
depend on all details of the model; however, such non-universality
is all contained in the function $K(\l)$ (which is expressed in
our analysis as a convergent power series expansion), and the
indices have a simple universal expressions in terms of the
parameter $K$.

From \pref{fon5} we see that, analogously to what happens for the critical
exponents, the {\it amplitude} of the dominant part, for $\pp\to 0$, of the
density-density correlation Fourier transform verifies an {\it universal}
relation in terms of $K$ and $v_s$; on the contrary no universal relation is
expected to be true for the amplitude of the Fourier transform close to $(\pm 2
p_F,0)$.

The equation \pref{gio} is an universal relation connecting the susceptibility
defined in \pref{kk} with $K$ and $v_s$; it is one of the two relations
conjectured by Haldane in \cite{[6]} (see (3) of \cite{[6]}, where $v_N\equiv
(\pi \k)^{-1}$ and $K\equiv e^{2\phi}$). Note that in the case of the $XYZ$
model ($J_1\not= J_2$) with $h=0$ the exponent $\bar\n$ has been computed by
Baxter and it has been found, see (10.12.24) of \cite{[15]}, if $\cos
\bar\m=-J_3/J_1=\l$,
\be\lb{bb1} \bar\n={\pi\over 2\bar\m}=1+{2\l\over\pi}+ O(\l^2)\;. \ee
From \pref{xxx} $K^{-1}=e^{-2\phi}=2(1-{\bar\m\over\pi})$. Moreover from the
Bethe ansatz solution \cite{[9]} exact expressions for $v_s$ and $\k$ can be
obtained,
%
\be\lb{bb2} v_s={\pi\over\bar\m}\sin\bar\m\quad\quad
\k= [2\pi(\pi/\bar\m-1)\sin\bar\m]^{-1}\;,\ee
so that \pref{gio} is verified. In general $\k,K,v_s$  depend on the magnetic
field $h$ and the specific form of the interaction $\hat v(k)$ (such dependence
is simple at first order, see \pref{l}, but in general quite complex), but our
theorem shows that the Kadanoff and Haldane relations \pref{xxx} and \pref{gio}
are still true. This is the first example in which such relations are proven in
generic non-solvable models.

In \cite{[14]} a statement similar to \pref{xxx}, \pref{xxxx} has
been proved in the case of planar spin models; the extension to
the present case is straightforward. The main novelty of the paper
is the proof of the Haldane relation \pref{gio}, so we will focus
on its derivation. The main ideas of our proof should be
understood also from people who did not read our previous papers,
only referring to them for the proof of several technical results
that we need.

\section{Proof of Theorem \ref{thm1}}

As the interaction modifies the value of the Fermi momentum and of the sound
velocity, it is convenient to include some part of the free hamiltonian in the
interaction part, by writing \pref{z} in the following way
\bea\lb{z1} &&H = H_0+ \n \sum_{x=1}^L a^+_{x} a^-_{x} -\d
\sum_{x=1}^L [\cos p_F a_{x}^+ a_{x}^-
-(a_{x+1}^+a_{x}^- + a_x^+ a_{x+1}^-)/2]\nn\\
&&+\l \sum_{1\le x,y\le L} v(x-y) a^+_x a^-_x a^+_{y}
a^-_{y},
\eea
with
\be
H_0=-{v_s\over v_F}\sum_{x=1}^L \fra12 [a^+_{x} a^-_{x+1}+ a_{x+1}^+ a^-_{x}-2
\cos p_F a^+_x a^-_x]
\ee
and
\be\lb{xc}
\cos p_F=-\l-h-\n\virg v_s=v_F(1+\d)\;.
\ee
Note that, if $H=H_0$, the Fourier transform of the 2-point function is
singular at $\kk=(\pm p_F,0)$ and the sound velocity is $v_s$. The parameter
$\n$ is chosen as a function of $\l$ and $p_F$, so that the singularity of the
Fourier transform of the two-point function corresponding to $H$ is fixed at
$\kk=(\pm p_F,0)$; the first equation in \pref{xc} gives the value of $h$
corresponding, in the model \pref{z}, to the chosen value of $p_F$. On the
contrary, the parameter $\d$ is an unknown function of $\l$ and $p_F$, whose
value is determined by requiring that, in the renormalization group analysis,
the corresponding marginal term flows to $0$; this implies that $v_s$ is the
sound velocity even for the full Hamiltonian $H$.

It is well known that the correlations of the quantum spin chain
can be derived by the following Grassmann integral, see \cite{[13]}:
\be\lb{1z}
e^{\WW_M(J,\tilde J,\phi)}=\int P(d\psi) e^{-\VV(\psi)+ \int d\xx [J_\xx \r_\xx +
\tilde J_\xx j_\xx] + \int d\xx [\phi^+_\xx \psi^-_{\xx} + \psi^-_{\xx}
\psi^+_\xx]}\;,
\ee
where $\psi^\pm_\xx$ and $\phi^\pm_\xx$ are Grassmann variables, $J_\xx$ and
$\tilde J_\xx$ are commuting variables, $\int d\xx$ is a shortcut for
$\sum_{x\in\L}\int_{-\b/2}^{\b/2} dx_0$, $P(d\psi)$ is a Grassmann Gaussian
measure in the field variables $\psi^\pm_\xx$ with covariance (the free
propagator) given by
\be
g_M(\xx-\yy)= {1\over\b L}\sum_{\kk\in\DD_{L,\b}}  {\chi(\g^{-M} k_0) e^{i\d_M
k_0} e^{i\kk(\xx-\yy)} \over -i k_0+ (v_s/v_F) (\cos p_F -\cos k)}\;,
\ee
where $\chi(t)$ is a smooth compact support function equal to $0$ if $|t|\ge
\g>1$ and equal to $1$ for $|t|<1$, $\kk=(k,k_0)$, $\kk\cdot\xx=k_0x_0+kx$,
${\cal D}_{L,\b}\={\cal D}_L \times {\cal D}_\b$, ${\cal D}_L\=\{k={2\pi n/L},
n\in \zzzz, -[L/2]\le n \le [(L-1)/2]\}$, ${\cal D}_\b\=\{k_0=2(n+1/2)\pi/\b,
n\in Z\}$ and
\bea
&&\VV(\psi)=\l \int d\xx d\yy \tilde v(\xx-\yy) \psi_\xx^+ \psi_\yy^+
\psi_\yy^- \psi_\xx^-+ \n \int d\xx \psi_{\xx}^+\psi_{\xx}^- -\nn\\
&&- \d \int d\xx [\cos p_F \psi_{\xx}^+ \psi_{\xx}^-
-(\psi_{\xx+\e_1}^+\psi_{\xx}^- + \psi_\xx^+ \psi_{\xx+\e_1}^-)/2]\;,\nn \eea
with $\e_1=(1,0)$, $\tilde v(\xx-\yy)=\d(x_0-y_0)v(x-y)$.
Moreover
\be
\r_\xx=\psi^+_{\xx}\psi^-_{\xx}\virg j_\xx= (2i
v_F)^{-1}[\psi_{\xx+\e_1}^+\psi_{\xx}^- -\psi_\xx^+ \psi_{\xx+\e_1}^-]\;.
\ee
Note that, due to the presence of the ultraviolet cut-off $\g^M$, the Grassmann
integral has a finite number of degree of freedom, hence it is well defined.
The constant $\d_M=\b/\sqrt{M}$ is introduced in order to take correctly into
account the discontinuity of the free propagator $g(\xx)$ at $\xx=0$, where it
has to be defined as $\lim_{x_0\to 0^-} g(0,x_0)$; in fact our definition
guarantees that $\lim_{M\to\io} g_M(\xx)=g(\xx)$ for $\xx\not=0$, while
$\lim_{M\to\io} g_M(0,0)=g(0,0^-)$.

We shall use the following definitions:
\bea\lb{corrXYZ}
&&G^{2,1}_\r(\xx,\yy,\zz) =\lim_{-l,N\to\io} {\partial\over\partial J_\xx}
{\partial^2 \over \partial\phi^+_\yy \partial\phi^-_\zz}
\WW_M(J,\tilde J,\phi)|_{J=\tilde J=\phi=0}\;,\nn\\
&&G^{2,1}_j(\xx,\yy,\zz) =\lim_{-l,N\to\io} {\partial\over\partial \tilde
J_\xx} {\partial^2 \over \partial\phi^+_\yy \partial\phi^-_\zz}
\WW_M(J,\tilde J,\phi)|_{J=\tilde J=\phi=0}\;,\nn\\
&&G^{2}(\yy,\zz) = \lim_{-l,N\to\io} {\partial^2 \over \partial\phi^+_\yy
\partial\phi^-_\zz} \WW_M(J,\tilde J,\phi)|_{J=\tilde J=\phi=0}\;,\\
&&G^{0,2}_{\r,\r}(\xx,\yy)=\lim_{-l,N\to\io}{\partial^2\over \dpr J_\xx \dpr
J_\yy}\WW_M(J,\tilde J,\phi)|_{J=\tilde J=\phi=0}\;.\nn
\eea
The Fourier transforms $\hat G^2(\kk)$ and $\hat G^{0,2}_{\r,\r}(\pp)$ of
$G^{2}(\yy,\zz)$ and $G^{2,0}_{\r,\r}(\xx,\yy)$ are defined in a way analogous
to the definition of $\hat\O(\pp)$ in \pref{FTO}. Moreover, we define the
Fourier transforms of $G^{2,1}_\a$, $\a=\r,j$, so that
\be\lb{FTG}
G^{2,1}_\a(\xx,\yy,\zz) = {1\over L\b} \sum_{\kk,\pp} e^{i\pp\xx -i(\kk+\pp)\yy
+i\kk\zz} \hat G^{2,1}_\a(\kk,\kk+\pp)\;.
\ee

The Grassmann integral \pref{1z} has been analyzed in \cite{[13],[13a]} by
Renormalization Group methods; by choosing properly the counterterms $\n$ and
$\d$, one gets expression which are uniformly analytic in $\b,L,M$. The
correlations obtained from the Grassmann integral coincide with the
correlations of the Hamiltonian model \pref{z1} as $M\to\io$. By such analysis
the asymptotic expressions \pref{a1} and \pref{a2} are proved, and the critical
indices $\eta$, $X_+$, $X_-$, and $\bar\n$ can be represented as power series
in the variable $r=\l_{-\io}/v_s$, where $\l_{-\io}=\l+O(\l^2)$ is the {\it
asymptotic effective coupling}. Such series are {\it convergent} for $r$ small
enough and their coefficients are {\it universal}, that is model independent.
Moreover, $v_s$ and $\l_{-\io}$ can be represented as power series of $\l$,
convergent near $\l=0$ and depending on all details of the model, so that this
property is true also for the critical indices. The fact that the critical
indices can be represented as universal functions of a single parameter implies
that they can be all expressed in terms of only one of them; however, to
compute explicitly such relations, by only using the complicated expansions in
terms of $r$, looks impossible.

The key observation is to take advantage from the gauge symmetries present in
the theory in the formal scaling limit. We introduce a continuum fermion model,
essentially coinciding with the formal scaling limit of the fermion model with
hamiltonian \pref{z1} (which is a QFT model), regularized by a non local fixed
interaction, together with an infrared $\g^l$ and ultraviolet $\g^N$ momentum
cut-offs, $-l,N\gg 0$. The limit $N\to\io$, followed from the limit $l\to-\io$,
will be called the {\it limit of removed cut-offs}. The model is expressed in
terms of the following Grassmann integral:
\bea\lb{vv1}
e^{\WW_{l,N}(J,\tilde J,\phi)} &=& \int\! P_Z(d\psi^{[l, N]})
e^{-\VV^{(N)}(\sqrt{Z}\psi^{[l ,N]}) + \sum_{\o=\pm} \int\! d\xx [Z^{(3)}
J_{\xx} +\o\, \tilde Z^{(3)} \tilde J_{\xx}] \r_{\xx,\o}}\cdot\nn\\
&\cdot& e^{Z \sum_{\o=\pm} \int d\xx [\psi^{+[l,N]}_{\xx,\o} \phi^-_{\xx,\o} +
\phi^+_{\xx,\o} \psi^{[l ,N]}]}\;,
\eea
where
\be\lb{rhodef}
\r_{\xx,\o} = \psi^{[l,N]+}_{\xx,\o} \ps^{[l,N]-}_{\xx,\o}\;,
\ee
$\xx\in\tilde\L$ and $\tilde\L$ is a square subset of $\RRR^2$ of size
$\g^{-l}$, say $\g^{-l}/2 \le |\tilde\L|\le \g^{-l}$, $P_Z(d\psi^{[l, N]})$ is
the fermionic measure with propagator
\be\lb{gth}
{1\over Z} g^{[l,N]}_{th,\o}(\xx-\yy)={1\over Z}{1\over
L^2}\sum_{\kk}e^{i\kk\xx}{\chi_{l,N}(\kk)\over -ik_0+\o c k}\;,
\ee
where $Z$ and $c$ are two parameters, to be fixed later, and $\chi_{l,N}(\kk)$
is the cutoff function. Moreover, the interaction is
\be\lb{gjhfk} \VV^{(N)}(\psi)={\l_\io\over 2} \sum_{\o}\int
d\xx \int d\yy v_0(\xx-\yy) \psi^+_{\xx,\o}
\psi^-_{\xx,\o}\psi^+_{\yy,-\o}\psi^-_{\yy,-\o}\;, \ee
where $v_0(\xx-\yy)$ is a rotational invariant potential, of the form
\be v_0(\xx-\yy)={1\over L^2}\sum_{\pp} \hat v_0(\pp)
e^{i\pp(\xx-\yy)}\;, \ee
with $|\hat v_0(\pp)|\le C e^{-\m |\pp|}$, for some constants $C$, $\m$, and
$\hat v_0(0)=1$. We shall use the following definitions, analogous to the
definitions \pref{corrXYZ} of the quantum spin chain:
\bea
&&G^{2,1}_{th,\r;\o}(\xx,\yy,\zz) =\lim_{-l,N\to\io} {\partial\over\partial
J_\xx} {\partial^2\over \partial\phi^+_{\yy,\o} \partial\phi^-_{\zz,\o}}
\WW_{l,N}(J,\tilde J,\phi)|_{J=\tilde J=\phi=0}\;,\nn\\
&&G^{2,1}_{th,j;\o}(\xx,\yy,\zz) =\lim_{-l,N\to\io} {\partial\over\partial
\tilde J_\xx} {\partial^2\over \partial\phi^+_{\yy,\o} \partial\phi^-_{\zz,\o}}
\WW_{l,N}(J,\tilde J,\phi)|_{J=\tilde J=\phi=0}\;,\nn\\
&&G^{2}_{th;\o}(\yy,\zz) = \lim_{-l,N\to\io} {\partial^2\over
\partial\phi^+_{\yy,\o} \partial\phi^-_{\zz,\o}} \WW_{l,N}(J,\tilde
J,\phi)|_{J=\tilde J=\phi=0}\;,\\
&&G^{0,2}_{th,\r,\r}(\xx,\yy)=\lim_{-l,N\to\io}{\partial^2\over \dpr J_\xx \dpr
J_\yy}\WW_{l,N}(J,\tilde J,\phi)|_{J=\tilde J=\phi=0}\;.\nn
\eea
The Fourier transforms $\hat G^2_{th;\o}(\kk)$ and $\hat
G^{0,2}_{th,\r,\r}(\pp)$ of $G^{2}_{th;\o}(\yy,\zz)$ and
$G^{2,0}_{th,\r,\r}(\xx,\yy)$ are defined in a way analogous to the definition
of $\hat\O(\pp)$ in \pref{FTO}. Moreover, we define the Fourier transforms of
$G^{2,1}_{th,\a;\o}$, $\a=\r,j$, as in \pref{FTG}.

In \S 3 of \cite{[17]} (see also \S 4 of \cite{[14]}) it has been proved that,
for small $\tilde\l_{\io}$ and for {\it non-exceptional momenta} (that is
$\kk$, $\pp$ and $\kk-\pp$ different from $0$),
\bea\lb{h11}
&&Z[-i p_0 {1\over Z^{(3)}} \hat G^{2,1}_{th,\r;\o}(\kk,\kk+\pp)+ \o p\ c
{1\over \tilde Z^{(3)}} \hat G^{2,1}_{th,j;\o}(\kk,\kk+\pp)]=\nn\\
&&\hspace{1cm} = A [\hat G^{2}_{th;\o}(\kk) - \hat G^{2}_{th;\o}(\kk+\pp)]\;,\\
&&Z [-i p_0 {1\over \tilde Z^{(3)}} \hat G^{2,1}_{th,j;\o}(\kk, \kk+\pp) + \o
p\ c {1\over Z^{(3)}} \hat G^{2,1}_{th,\r;\o}(\kk,\kk+\pp)] =\nn\\
&&\hspace{1cm} = \o\bar A [\hat G^{2}_{th;\o}(\kk) - \hat
G^{2}_{th;\o}(\kk+\pp)]\;,\nn
\eea
with
\be\lb{v2} A^{-1}=1-\t \virg \bar
A^{-1}=1+\t \virg \t={\l_\io\over 4\pi c}\;.
\ee
Equations \pref{h11} are the Ward Identities associated to the invariance of
the formal lagrangian with respect to local and local chiral Gauge
transformations. The fact that $A,\bar A$ are not equal to $1$ is a well known
manifestation of the {\it anomalies} in quantum field theory; naively, by a
gauge transformation in the non regularized ill defined Grassmann integrals,
one would get similar expressions with $A=\bar A=1$. Finally, the linearity of
$A^{-1},\bar A^{-1}$ in terms of $\l_{\io}$ is a property called {\it anomaly
non-renormalization} and it depends crucially on the regularizations used; with
different regularizations such a property could be violated, see \cite{[16]}.

An easy extension of the results given in \cite{[17]} allows us to deduce also
a set of Ward Identities for the continuum model correlations of the {\it
density operator} $\r_{\xx,\o}$ defined in \pref{rhodef}. To be more precise,
let us consider the functional
\be\lb{vv1a}
e^{\tilde\WW(J)} = \int\! P_Z(d\psi) e^{-\VV^{(N)}(\sqrt{Z}\psi) + \sum_{\o}\int\!
d\xx J_{\xx,\o} \r_{\xx,\o}}\;,
\ee
and let us define
\be
G_{\o,\o'}(\xx,\yy)= \lim_{-l,N\to\io}{\partial^2\over \dpr J_{\xx,\o} \dpr
J_{\yy,\o'}} \tilde \WW(J)|_{J=0}\;.
\ee
In App. \ref{app1} we shall prove that, in the limit $-l,N\to\io$,
\bea\lb{tt}
&&D_\o(\pp) \hat G_{\o,\o}(\pp) - \t\ \hat v_0(\pp) D_{-\o}(\pp) \hat
G_{-\o,\o}(\pp) + {1\over 4\pi c Z^2} D_{-\o}(\pp)=0\;,\nn\\
&&D_{-\o}(\pp) \hat G_{-\o,\o}(\pp) - \t\  \hat v_0(\pp) D_{\o}(\pp) \hat
G_{\o,\o}(\pp) =0\;,
\eea
where
\be D_\o(\pp)=-i p_0+\o c p\;. \ee

By using \pref{tt} and $\hat v_0(\pp) = 1 +O(\pp)$, we get:
\bea
&&\hat G_{\o,\o}(\pp) = -{1\over Z^2}{1\over 4\p c (1-\t^2)}
{D_{-\o}(\pp)\over D_\o(\pp)}+O(\pp)\;,\nn\\
&&\hat G_{-\o,\o}(\pp) = -{1\over Z^2}{\t\over 4\pi c(1-\t^2)} +O(\pp)\;,
\eea
which implies, after a few simple calculations, that
\be\lb{ggv1}
\hat G^{0,2}_{th,\r,\r}= -{1\over 4\pi c Z^2}{(Z^{(3)})^2\over 1-\t^2}
\left[{D_-(\pp)\over D_+(\pp)}+{D_+(\pp)\over D_-(\pp)}+2 \t \right] +O(\pp)\;.
\ee

The crucial point is that it is possible to choose the parameters of the
continuum model so that the correlations in the two models are the same, up to
small corrections, for small momenta.

\begin{lemma}\lb{lm1} Given $\l$ small enough, there are constants $Z$, $Z^{(3)}$,
$\tilde Z^{(3)}$, $\l_\io$, depending analytically on $\l$, such that, if we
put $c=v_s$, the critical indices of the two models coincide. Moreover, if $\k
\le 1$ and $|\pp|\le \k$,
\be\lb{h10}
\hat G^{0,2}_{\r,\r}(\pp) = \hat G^{0,2}_{th,\r,\r}(\pp)+A_{\r,\r}(\pp)\;,
\ee
with $A_{\r,\r}(\pp)$ continuous in $\pp$ and $O(\l)$. Finally, if we put
$\pp_F^\o=(0,\o p_F)$ and we suppose that $0<\k\le |\pp|,|\kk'|,|\kk'-\pp|\le
2\k$, $0<\th<1$, then
\bea\lb{h10a}
&&\hat G^{2,1}_\r(\kk'+ \pp_F^\o, \kk'+\pp+\pp_F^\o)
= \hat G^{2,1}_{th,\r;\o}(\kk',\kk'+\pp)[1+O(\k^\th)]\;,\nn\\
&&\hat G^{2,1}_j(\kk'+ \pp_F^\o, \kk'+\pp+\pp_F^\o)
= \hat G^{2,1}_{th,j;\o}(\kk',\kk'+\pp)[1+O(\k^\th)]\;,\\
&&\hat G^2(\kk'+\pp_F^\o) = \hat G^{2}_{th,\o}(\kk')[1+O(\k^\th)]\;.\nn
\eea
\end{lemma}

This Lemma will be proved in the next section; we now exploit its implications.

\vspace{.3cm}

By combining \pref{h10a} and \pref{h11} we find that
\bea\lb{xa1}
&&-i p_0\ \hat G^{2,1}_\r(\kk'+ \pp_F^\o, \kk'+\pp+\pp_F^\o) + \o p\ \tilde v_J
\hat G^{2,1}_j(\kk'+ \pp_F^\o, \kk'+\pp+\pp_F^\o) =\nn\\
&&={Z^{(3)}\over (1-\t)Z} \left[ \hat G^2(\kk'+\pp_F^\o) - \hat
G^2(\kk'+\pp+\pp_F^\o) \right ] [1+O(\k^\th)]
\eea
and
\bea\lb{xa3}
&&-i p_0\ \hat G^{2,1}_j(\kk'+ \pp_F^\o, \kk'+\pp+\pp_F^\o) + \o p\ \tilde v_N
\hat G^{2,1}_\r(\kk'+ \pp_F^\o, \kk'+\pp+\pp_F^\o) =\nn\\
&&={\tilde Z^{(3)}\over (1+\t)Z} \left[ \hat G^2(\kk'+\pp_F^\o) - \hat
G^2(\kk'+\pp+\pp_F^\o) \right ] [1+O(\k^\th)]\;,
\eea
with
\be\lb{bert} \tilde v_N=v_s {Z^{(3)}\over \tilde Z^{(3)}}\virg \tilde v_J=v_s
{\tilde Z^{(3)}\over Z^{(3)}}\;.
\ee
On the other hand, a WI for the model \pref{z} can be derived directly from the
commutation relations, see App. \ref{app2}; one gets
\bea\lb{xa1a}
&&-i p_0\ \hat G^{2,1}_\r(\kk'+ \pp_F^\o, \kk'+\pp+\pp_F^\o) + \o p\ v_F
\hat G^{2,1}_j(\kk'+ \pp_F^\o, \kk'+\pp+\pp_F^\o) =\nn\\
&&= \left[ \hat G^2(\kk'+\pp_F^\o) - \hat G^2(\kk'+\pp+\pp_F^\o) \right ]
[1+O(\k^\th)]\;.
\eea
Hence, if we compare \pref{xa1a} with \pref{xa1}, we get the identities
\be\lb{hh} {Z^{(3)}\over (1-\t)Z}=1\virg \tilde v_J=v_F\;.\ee
Moreover, in App. \ref{app2} we also show that
\be\lb{jjj1}
\hat G^{0,2}_{\r,\r}(\pp) =0 \virg \hbox{if\ } \pp=(0,p_0)\;,
\ee
and this fixes the value of $A_{\r\r}(0)$ so that
\bea\lb{berc}
&&\hspace{-0.5cm}\hat G^{0,2}_{\r,\r}(\pp) = {1\over 4\pi v_s
Z^2}{(Z^{(3)})^2\over 1-(\l_\io/4\p v_s)^2} \left[2-{D_-(\pp)\over
D_+(\pp)}-{D_+(\pp)\over D_-(\pp)} \right] + R(\pp)\;,
\eea
with $R({\bf 0})=0$.
%
%
By using \pref{hh}, we get \pref{fon5}, with
\be\lb{ggha}
K= {1\over Z^2}{(Z^{(3)})^2\over 1-(\l_\io/4\p v_s)^2} = {1-(\l_\io/4\p v_s)
\over 1+(\l_\io/4\p v_s)}\;.
\ee

It has been proved in Theorem 4.1 of \cite{[14]} (where we used $c=1$) that the
critical indices of the model \pref{vv1} have a simple expressions in terms of
$\l_{\io}$; if we take eq. (4.26) of \cite{[14]} and we put $\t=\l_\io/4\p
v_s$, we get:
\be X_+=1-{(\l_\io/2\p v_s)\over 1+(\l_\io/4\p
v_s)}\virg X_-=1+{(\l_\io/2\p v_s)\over 1-(\l_\io/4\p v_s)}\;; \ee
this implies the relations \pref{xxx}, with $K$ given by \pref{ggha}. Eq.
\pref{l} follows from the remark that, at the first order, $\l_\io=\l_{-\io}$,
while $\l_{-\io}$, which was imposed to be equal in the two models, is related
to $\l$ (always at the first order) by the relation $\l_{-\io}=2\l [\hat v(0) -
\hat v(2 p_F)]$. The first identity in \pref{xxxx} is proved as eq. (1.11) of
\cite{[14]}; note that $\bar \n$ is different from the index $\n$ appearing in
\cite{[14]}, but one can see that this difference only implies that one has to
replace, in eq. (1.11) of \cite{[14]}, $x_+$ with $x_-$. Finally, by using the
identity (4.21) of \cite{[14]} (where $\h$ is denoted $\h_z$), we get also the
second identity in \pref{xxxx}. The proof of Theorem \ref{thm1} is completed.

\vspace{.3cm}

{\bf Remark 1 - } Note that in the WI \pref{xa1}, \pref{xa3} for the model
\pref{z1} {\it three} different velocities appear. This is due to the fact that
the {\it irrelevant} operators (in the RG sense) break the relativistic
symmetries present in the model in the scaling limit and produce different
renormalization of the velocities. Note also that the velocities $\tilde
v_N,\tilde v_J$ defined in \pref{bert} verify the universal relation
\be\lb{b2}
\tilde v_N \tilde v_J=v_s^2\;.
\ee

\vspace{.3cm}

{\bf Remark 2 - } The constraints \pref{hh} and \pref{jjj1} on the
renormalization parameters of the continuum model, which describes the large
distance behavior, are a consequence of the existence of a well defined lattice
hamiltonian.

\section{Proof of Lemma \ref{lm1}}

The proof of the lemma is based on the RG analysis of the Grassmann integrals
\pref{1z} and \pref{vv1}, described in \cite{[13],[13a]} and \cite{[16],[17]},
respectively.

Let us recall briefly the analysis of \ref{1z}. Let $T^1$ be the one
dimensional torus, $||k-k'||_{T^1}$ the usual distance between $k$ and $k'$ in
$T^1$ and $||k||=||k-0||$. We introduce a {\sl scaling parameter} $\g>1$ and a
positive function $\c(\kk') \in C^{\io}(T^1\times R)$, $\kk'=(k',k_0)$, such
that $ \c(\kk') = \c(-\kk') = 1$ if $|\kk'| <t_0 = a_0 v_s/\g$ and $=0$ if
$|\kk'|
>a_0$ where $a_0= \min \{{p_F\over 2}, {\p- p_F\over 2}\}$ and
$|\kk'|=\sqrt{k_0^2+(v_s ||k'||_{T^1})^2}$. The above definition
is such that the supports of $\c(k-p_F,k_0)$ and $\c(k+p_F,k_0)$
are disjoint and the $C^\io$ function on $T^1\times R$
\be
\hat f_1(\kk) \= 1- \c(k-p_F,k_0) - \c(k+p_F,k_0)
\ee
is equal  to $0$, if $[||v_s (|k|-p_F)||_{T^1}]^2 +k_0^2<t_0^2$.

We define also, for any integer $h\le 0$,
\be
f_h(\kk')= \c(\g^{-h}\kk')-\c(\g^{-h+1}\kk')\;.
\ee
We have
\be
\c(\kk') = \sum_{h=h_{L,\b}}^0 f_h(\kk')\;,
\ee
where
\be
h_{L,\b} =\min \{h:t_0\g^{h+1} > \sqrt{(\p\b^{-1})^2+(v_s\p L^{-1})^2} \}\;.
\ee
Note that, if $h\le 0$, $f_h(\kk') = 0$ for $|\kk'| <t_0\g^{h-1}$ or $|\kk'|
>t_0 \g^{h+1}$, and $f_h(\kk')= 1$, if $|\kk'| =t_0\g^h$. Let us now define:
\be
\hat f_h(\kk) = f_h(k-p_F,k_0) +f_h(k+p_F,k_0)\;.
\ee
This definition implies that, if $h\le 0$, the support of $\hat f_h(\kk)$ is
the union of two disjoint sets, $A_h^+$ and $A_h^-$. In $A_h^+$, $k$ is
strictly positive and $||k-p_F||_{T^1}\le t_0\g^h \le t_0$, while, in $A_h^-$,
$k$ is strictly negative and $||k+p_F||_{T^1}\le t_0\g^h$. The label $h$ is
called the {\sl scale} or {\sl frequency} label. Note that
\be
1=\sum_{h=h_{L,\b}}^1 \hat f_h(\kk)\;;
\ee
hence, if we approximate $p_F$ by $(2\p/L)(n_F+1/2)$, $n_F$ equal to the
integer part of $Lp_F/(2\p)$, and we define $\DD'_L= \{k'=2(n+1/2)\pi/L, n\in
\ZZZ, -[L/2]\le n \le [(L-1)/2]\}$ and $\DD'_{L,\b}=\DD'_L \times \DD_\b$, we
can write:
\bea\lb{gdef}
&&g(\xx-\yy)= g^{(1)}(\xx-\yy) + \sum_{\o=\pm} \sum_{h=h_{L,\b}}^0
e^{-ip_F(x-y)} g^{(h)}_\o(\xx-\yy)\;,\nn\\
&&g^{(1)}(\xx-\yy) = {1\over\b L} \sum_{\kk\in\DD_{L,\b}} e^{-i\kk(\xx-\yy)}
{\hat f_1(\kk)\over -i k_0 + (v_s/v_F)(\cos p_F -\cos k)}\;,\\
&&g^{(h)}_\o(\xx-\yy) = {1\over\b L} \sum_{\kk'\in\DD'_{L,\b}}
e^{-i\kk'(\xx-\yy)} {f_h(\kk')\over -i k_0+ E_\o(k')}\;,\nn
\eea
where
\be
E_\o(k') = \o v_s\sin k' + (1+\d)\cos p_F (1-\cos k')\;.
\ee

Let us now describe the perturbative expansion of $\WW$; for simplicity we
shall consider only the case $\phi=0$. We can write:
\bea\lb{2z}
&& e^{\WW(J,\tilde J,0)}= \int P(d\psi^{\le 0}) \int P(d\psi^{(1)})
e^{-\VV(\psi) + \int d\xx [J_\xx \r_\xx+\tilde J_\xx j_\xx]}=\nn\\
&&=e^{-L\b E_0} \int P(d\psi^{\le 0})e^{-\VV^{(0)}(\psi^{\le
0})+\BB^{(0)}(\psi^{\le 0},J,\tilde J)}\;,
\eea
where, if we put $\ux=(\xx_1,\ldots,\xx_{2n})$, $\oo=(\o_1,\ldots,\o_{2n})$ and
$\psi_{\ux,\oo} = \prod_{i=1}^n$ $\psi^+_{\xx_i,\o_i} \prod_{i=n+1}^{2n}
\psi^-_{\xx_i,\o_i}$, the {\it effective potential} $\tilde\VV^{(0)}(\ps)$ can
be represented as
\be\lb{3.2aaa}
\VV^{(0)}(\psi)= \sum_{n\ge 1} \sum_{\oo} \int d\ux W^{(0)}_{\oo, 2n}(\ux)
\psi_{\ux,\oo}\;,
\ee
the kernels $W^{(0)}_{\oo,2n}(\ux)$ being analytic functions of $\l$ and $\n$
near the origin; if $|\n|\le C|\l|$ and we put $\uk=(\kk_1, \ldots,
\kk_{2n-1})$, their Fourier transforms satisfy, for any $n\ge 1$, the bounds,
see \S 2.4 of \cite{[13]},
\be |\widehat W^{(0)}_{\oo,2n}(\uk)| \le C^n |\l|^{\max\{1,n-1\}}\;.
\ee
A similar representation can be written for the functional $\BB^{(0)}(\psi^{\le
0},J,\tilde J)$, containing all terms which are at least of order one in the
external fields, including those which are independent on $\psi^{\le 0}$.

The integration of the scales $h\le 0$ is done iteratively in the following
way. Suppose that we have integrated the scale $0,-1,-2,..,j$, obtaining
\be\lb{61}
e^{\WW(J,\tilde J,0)}=e^{-L\b E_j} \int P_{Z_j,C_j}(d\psi^{\le
j})e^{-\VV^{(j)}(\sqrt{Z_j}\psi^{\le j})+\BB^{(j)}(\sqrt{Z_j}\psi^{\le
j},J,\tilde J)}\;,
\ee
where, if we put $C_j(\kk')^{-1}=\sum_{h=h_{L,\b}}^j f_h(\kk')$, $P_{Z_j,C_j}$
is the Grassmann integration with propagator
\be\lb{62}
{1\over Z_j}\, g^{(\le j)}_\o(\xx-\yy)= {1\over Z_j}{1\over\b L}
\sum_{\kk\in\DD'_{L,\b}} e^{i\kk(\xx-\yy)}{C_j^{-1}(\kk) \over -i
k_0+E_\o(k')}\;,
\ee
$\VV^{(j)}(\psi)$ is of the form
\be\lb{3.2aaax}
\VV^{(j)}(\psi)= \sum_{n\ge 1} \sum_{\oo} \int d\ux W^{(j)}_{\oo,2n}(\ux)
\psi_{\ux,\oo}\;,
\ee
and $\BB^{(j)}(\psi^{\le j},J,\tilde J)$ contains all terms which are at least
of order one in the external fields, including those which are independent on
$\psi^{\le j}$. For $j=0$, $Z_0=1$ and the functional $\VV^{(0)}$ and
$\BB^{(0)}$ are exactly those appearing in \pref{2z}.

First of all, we define a localization operator (see \cite{[13],[13a]} for
details) in the following way:
\bea\lb{h1}
&&\hspace{-.3cm}\LL \VV^{(j)}(\sqrt{Z_j}\psi)=\g^j n_j {Z_j\over \b
L}\sum_{\kk} \psi^+_{\kk,\o}\psi^-_{\kk,\o}+ a_j {Z_j\over \b L}\sum_{\kk}
E_\o(\kk) \psi^+_{\kk,\o}\psi^-_{\kk,\o}+\\
&&\hspace{-.3cm} z_j {Z_j\over \b L}\sum_{\kk} (-i k_0)
\psi^+_{\kk,\o}\psi^-_{\kk,\o} + l_j {Z_j^2\over (\b L)^4}\sum_{\kk_1,\kk',\pp}
\psi^+_{\kk,+}\psi^-_{\kk-\pp,+}\psi^+_{\kk',-}\psi^-_{\kk'+\pp,-}\;,\nn
\eea
\bea\lb{h2}
&&\hspace{.5cm} \LL \BB^{(j)}(\sqrt{Z_j}\psi)={Z^{(1)}_j\over (\b L)^2}
\sum_{\kk,\pp} J_\pp \Big[\sum_\o \psi^+_{\kk,\o}\psi^-_{\kk-\pp,\o}\Big]+\\
&&+{Z^{(2)}_j\over (\b L)^2} \sum_{\kk,\pp} J_{\pp+2\o\pp_F} \Big[\sum_\o
\psi^+_{\kk,\o} \psi^-_{\kk-\pp,-\o}\Big] + {\tilde Z^{(1)}_j\over (\b L)^2}
\sum_{\kk,\pp} J_\pp \Big[\sum_\o \o\, \psi^+_{\kk,\o}
\psi^-_{\kk-\pp,\o}\nn\Big]\nn\\
&&\hspace{.5cm} +{\tilde Z^{(2)}_j\over (\b L)^2} \sum_{\kk,\pp} \tilde
J_{\pp+2\o\pp_F} \Big[\sum_\o
\o\,\psi^+_{\kk,\o}\psi^-_{\kk-\pp,-\o}\Big]\;,\nn
\eea
where $\pp_F=(p_F,0)$. This definitions are such that the difference between
$-\VV^{(j)} + \BB^{(j)}$ and $-\LL \VV^{(j)} + \LL \BB^{(j)}$ is made of
irrelevant terms.

The constants appearing in \pref{h1} and \pref{h2} are evaluated in terms of
the values of the corresponding kernels at zero external momenta. Since the
space momentum $k$ of $\psi^+_{\kk,\o}$ is measured from the Fermi surface,
this means that the external momenta corresponding to the fermion variables are
put equal to $(\o p_F,0)$, while $\pp$ is put equal to $(0,0)$. On the other
hand, it is easy to see that the kernel multiplying $J\psi^+\psi^-$ is {\it
even} in the exchange $k\to -k$ ($k$ is here the true space momentum, not the
momentum measured from the Fermi surface), since both the propagator and the
interaction are even, while the kernel multiplying $\tilde J\psi^+\psi^-$ is
{\it odd} in the exchange $k\to -k$, because of the parity properties of the
current $j_\xx$. These considerations are used in the definition of the
constants in \pref{h2}.

We then renormalize the integration measure, by moving to it some of the
quadratic terms in the r.h.s. of \pref{h1}, that is $z_j (\b L)^{-1} \sum_{\kk}
[-i k_0 + E_\o(\kk)] \psi^+_{\kk,\o}\psi^-_{\kk,\o}$; the Grassmann integral in
the r.h.s. of \pref{61} takes the form:
\be\lb{61a}
\int P_{\tilde Z_{j-1},C_j} (d\psi^{(\le j)}) e^{-\tilde\VV^{(j)}(\sqrt{Z_j}
\psi^{\le j}) + \BB^{(j)}(\sqrt{Z_j} \psi^{\le j},J,\tilde J)}\;,
\ee
where $\tilde\VV^{(j)}$ is the remaining part of the effective interaction and
$P_{\tilde Z_{j-1},C_j}(d\psi^{\le j})$ is the measure whose propagator is
obtained by substituting in \pref{62} $Z_j$ with
\be\tilde Z_{j-1}(\kk) =Z_j [1+z_j C_j(\kk)^{-1}]\;. \ee
It is easy to see that we can decompose the fermion field as $\psi^{\le j} =
\psi^{\le j-1} + \psi^{(j)}$, so that
\be
P_{\tilde Z_{j-1},C_j}(d\psi^{\le j}) = P_{ Z_{j-1}, C_{j-1}}(d\psi^{(\le
j-1)}) P_{ Z_{j-1}, \tilde f_j^{-1}}(d\psi^{(j)})\;,
\ee
where $\tilde f_j(\kk)$ (see eq. (2.90) of \cite{[13]}) has the same support
and scaling properties as $f_j(\kk)$. Hence, if make the field rescaling
$\psi\to [\sqrt{Z_{j-1}}/ \sqrt{Z_j}]\psi$ and we call
$\hat\VV^{(j)}(\sqrt{Z_{j-1}} \psi^{\le j})$ the new effective potential, we
can write \pref{61a} in the form
\bea
&&\int P_{Z_{j-1},C_{j-1}} (d\psi^{(\le j-1)}) \int P_{ Z_{j-1}, \tilde
f_j^{-1}}(d\psi^{(j)})\cdot\\
&&\cdot e^{-\hat\VV^{(j)}(\sqrt{Z_{j-1}} \psi^{\le j}) +
\hat\BB^{(j)}(\sqrt{Z_{j-1}} \psi^{\le j},J,\tilde J)}\;.\nn
\eea
By performing the integration over $\psi^{(j)}$, we finally get \pref{61}, with
$j-1$ in place of $j$.

In order to analyze the result of this iterative procedure, we note that $\LL
\hat\VV^{(j)}(\psi)$ can be written as
\be
\LL \hat\VV^{(j)}(\psi) = \g^j\n_j F_\n(\psi) + \d_j F_\a(\psi) + \l_j
F_\l(\psi)\;,
\ee
where $F_\n(\psi)$, $F_\a(\psi)$ and $F_\l(\psi)$ are the functions of $\psi$,
which appear in \pref{h1} in the terms proportional to $n_j$, $a_j$ and $l_j$,
respectively. $\n_j=(\sqrt{Z_j}/ \sqrt{Z_{j-1}}) n_j$, $\d_j=(\sqrt{Z_j}/
\sqrt{Z_{j-1}}) (a_j-z_j)$ and $\l_j=(\sqrt{Z_j}/ \sqrt{Z_{j-1}})^2 l_j$ are
called the {\it running couplings} (r.c.) on scale $j$. In Theorem (3.12) of
\cite{[13]} it is proved that the kernels of $\hat\VV^{(j)}$ and
$\hat\BB^{(j)}$ are {\it analytic} as functions of the r.c., provided that they
are small enough. One has then to analyze the flow of the r.c. (the {\it beta
function}) as $j\to-\io$. We shall now summarize the results, explained in
detail in \cite{[13],[13b]}.

The propagator $\tilde g^{(j)}_\o(\xx-\yy)$ of the single scale measure $P_{
Z_{j-1}, \tilde f_j^{-1}}$, can be decomposed as
\be\lb{ne1} \tilde g^{(j)}_\o(\xx-\yy)= {1\over Z_j}\,
g^{(j)}_{th,\o}(\xx-\yy)+r_j(\xx-\yy)\;, \ee
where
\be\lb{gjth} {1\over Z_j}\, g^{(j)}_{th,\o}(\xx-\yy)= {1\over Z_j}{1\over\b
L}\sum_{\kk\in\DD_{L,\b}}e^{i\kk(\xx-\yy)}{f_j(\kk) \over -i k_0+\o v_s k}
\ee
describes the leading asymptotic behavior, while the remainder $r_j(\xx-\yy)$
satisfies, for any $M>0$ and $\th<1$, the bound
\be |r_j(\xx-\yy)|\le {\g^{(1+\th)j}\over Z_j} {C_{M,\th}\over 1+(\g^j
|\xx-\yy|^M)}\;. \ee
We call $Z^{(th)}_j$ the values of $Z_j$ one would obtain by substituting
$\VV^{(0)}$ with $\LL \VV^{(0)}$ and by putting $r_h=0$ for any $h\ge j$ and we
observe that, by (4.50) of \cite{[13]},
\be\lb{ne}
\left| {Z_{j}\over Z_{j-1}}-{Z^{(th)}_{j}\over Z^{(th)}_{j-1}}\right|\le
C_\th\l^2\g^{\th j}\;.
\ee
\pref{ne1} and \pref{ne} imply that, see \S 4.6 of \cite{[13]}, the r.c.
satisfy recursive equations of the form:
\bea
&&\l_{j-1}=\l_j+\b_\l^{(j)}(\l_j,...,\l_0)+
\bar\b_\l^{(j)}(\l_j,\d_j,\n_j;...;\l_0,\d_0,\n_0)\;,\nn\\
&&\lb{bb} \d_{j-1}=\d_j+\b_\d^{(j)}(\l_j,\d_j,\n_j;...;\l_0,\d_0,\n_0)\;,\\
&&\n_{j-1}=\g\n_j+\b_\n^{(j)}(\l_j,\d_j,\n_j;...;\l_0,\d_0,\n_0)\;, \nn
\eea
where $\b_\l^{(j)}$, $\bar\b_\l^{(j)}$, $\b_\d^{(j)}$, $\b_\n^{(j)}$ can be
written as {\it convergent} expansions in their arguments, if $\e_j=\max_{j\le
h\le 0} \max \{|\l_h|,|\d_h|,|\n_h|\}$ is small enough. By definition,
$\b_\l^{(j)}$ is given by a sum of multiscale graphs (collected in trees; their
definition is in \S 3 of \cite{[13]}), containing only $\l$-vertices with scale
$\le 0$ and in which the propagators $g^{(h)}_{\o}$ and the wave function
renormalizations $Z_h$, $0\ge h\ge j$, are replaced by $g^{(h)}_{th,\o}$ and
$Z_h^{(th)}$, $0\ge h\ge j$; $\bar\b_\l^{(j)}$ contains the correction terms
together with the remainder of the expansion.

The following crucial property, called {\it vanishing of the Beta function},
was proved by means of Ward Identities in \cite{[13b]}; for any $\th<1$,
\be\lb{beta}
|\b_\l^{(j)}(\l_j,...,\l_j)|\le C_\th |\l_j|^2\g^{\th j}\;.
\ee
It is also possible to prove that, for a suitable choice of $\d,\n=O(\l)$,
$\d_j,\n_j=O(\g^{\th j}\bar\l_j)$, if $\bar\l_j=\sup_{k\ge j}|\l_k|$, and this
implies, by the {\it short memory property} (exponential decreasing
contribution of the graphs with propagators of scale $h>j$, as $h-j$ grows,
see the remark after (4.31) of \cite{[13]}),
that $\bar\b_\l^{(j)}=O(\g^{\th j}\bar\l_j^2)$, so that the sequence $\l_j$
converges, as $j\to -\io$, to a smooth function $\l_{-\io}(\l)= \l +O(\l^2)$,
such that
\be\lb{2.42a}
|\l_j- \l_{-\io}| \le C_\th \l^2 \g^{\th j}\;.
\ee
In a similar way we can also analyze the {\it renormalization constants}
$Z_j^{(\a)}$ and $\tilde Z_j^{(\a)}$, $\a=1,2$, defined in \pref{h2}, and the
field strength renormalization $Z_j$; we can write:
\be\lb{ffgz}
{Z_{j-1}\over Z_j} = 1+ \b_z^{(j)}(\l_j,...,\l_0)+
\bar\b_z^{(j)}(\l_j,\d_j;..,\l_0,\d_0)\;,
\ee
\be\lb{ffg}
{Z^{(\a)}_{j-1}\over Z^{(\a)}_j} = 1+ \b_{(\r,\a)}^{(j)}(\l_j,...,\l_0)+
\bar\b_{(\r,\a)}^{(j)}(\l_j,\d_j;..,\l_0,\d_0)\;,
\ee
\be\lb{ffga}
{\tilde Z^{(\a)}_{j-1}\over \tilde Z^{(\a)}_j} = 1+
\b_{(J,\a)}^{(j)}(\l_j,...,\l_0)+
\bar\b_{J,\a}^{(j)}(\l_j,\d_j;..,\l_0,\d_0)\;,
\ee
where, by definition, the $\b_t^{(j)}$ functions (with $t=z$, $(\r,\a)$ or
$(J,\a)$) are given by a sum of multiscale graphs, containing only
$\l$-vertices with scale $\le 0$ and in which the the propagators $g^{(h)}_{\o}$
and the renormalization constants $Z_h$, $Z^{(\a)}_h$, $\tilde Z^{(\a)}_h$,
$0\ge h\ge j$, are replaced by $g^{(h)}_{th,\o}$, $Z_h^{(th)}$, $Z^{(th,\a)}_h$
and $\tilde Z^{(th,\a)}_h$ (the definition of $Z^{(th,\a)}_h$, $\tilde
Z^{(th,\a)}_h$ is analogue to the one of $Z_h^{(th)}$); the $\bar\b_t^{(j)}$
functions contain the correction terms together the remainder of the expansion.
Note that, by definition, the constants $Z_j^{(th)}$ are exactly those
generated by \pref{ffgz} with $\bar\b_z^{(j)}=0$. Note that
$\bar\b_t^{(j)}=O(\l_j\g^{\th j})$ and, by using \pref{2.42a} and the short
memory property (see \eg \S 4.9 of \cite{[13]})
\be\lb{lau11}
\b^{(j)}_t(\l_j,...,\l_0)= \b^{(j)}_t(\l_{-\io},...,\l_{-\io})+O(\l\g^{\th
h})\;.
\ee
This implies that there exist, if $w$ is small enough, analytic functions
$\h_t(w)$, $t=z, (\r,\a), (J,\a)$, of order $\l^2$ for $t=z, (\r,1), (J,1)$ and
order $\l$ for $t=(\r,2), (J,2)$, such that
\bea\lb{lau12}
&&|\log_\g(Z_{j-1}/ Z_j) - \eta_z(\l_{-\io}/v_s)| \le
C_\th \l^2 \g^{\th j}\;,\nn\\
&&|\log_\g(Z^{(\a)}_{j-1}/ Z^{(\a)}_j) -
\eta_{\r,\a}(\l_{-\io}/v_s)| \le C_\th \l^2 \g^{\th j}\;,\\
&&|\log_\g( \tilde Z^{(\a)}_{j-1}/ \tilde Z^{(\a)}_j) -
\eta_{J,\a}(\l_{-\io}/v_s)| \le C_\th\l^2 \g^{\th j}\;.\nn
\eea
The fact that the {\it critical indices} $\h_t$ are functions of
$\l_{-\io}/v_s$ (not of $\l_{-\io}$ and $v_s$ separately) is not stressed in
\cite{[13],[13a]}, but follows very easily from dimensional arguments. It is
also easy to see that (see \cite{[13a]}, \S3.4), since the propagator
\pref{gjth} satisfies the symmetry property
\be\lb{symm}
\hat g^{(j)}_{th,\o}(k,k_0) = -i\o \hat g^{(j)}_{th,\o}(-k_0/v_s, v_s k)\;,
\ee
then $\eta_{\r,\a}(w)=\eta_{J,\a}(w)$, $\a=1,2$. Moreover, by using the
approximate Ward identities associated to the linearity in $\kk$ of $\hat
g^{(j)}_{th,\o}(\kk)^{-1}$, one can show (see Theorem 5.6 of \cite{[13]}) that
$\h_z=\h_{\r,1}$.

\*

The analysis of the functional \pref{vv1} can be done in a similar way. Even in
this case, we shall only sketch the main results, by referring to \cite{[17]}
and \cite{[14]} for more details. Again we perform a multiscale integration,
but now we have to consider two different regimes: the first regime, called
{\it ultraviolet}, contains the scales $0\le h\le N$, while the second one
contains the scales $h<0$, and is called {\it infrared}.

After the integration of the ultraviolet scales, see \cite{[17],[14]} (where
the external fields $J,\tilde J$ are substituted by two equivalent fields
$J_\o, \o=\pm1$), we can write the r.h.s. of \pref{vv1}, with $\phi=0$, as
\be\lb{221} \lim_{l\to -\io}\lim_{N \to \io} \int
P_Z(d\psi^{(\le 0)}) e^{-\bar\VV^{(0)}(\psi^{(\le 0)}) +
\bar\BB^{(0)}(\psi^{(\le 0)},J,\tilde J)}\;,
\ee
where the integration measure has a propagator $Z^{-1} g_{th,\o}^{(\le
0)}(\xx-\yy)$, given by \pref{gth} with $N=0$; moreover, $\bar\VV^{(0)}$ and
$\bar\BB^{(0)}$ are functionals similar to the functionals $\VV^{(0)}$ and
$\BB^{(0)}$ of \pref{2z}, with the following main differences. First of all,
$\LL \bar\VV^{(0)}$ can be written as in \pref{h1}, with  $E_\o(\kk)=c\,\o k$,
$n_0=0$, $a_0=z_0$ (these two properties easily from the symmetries of the
propagator) and $\l_0$ replaced by a new constant $\tilde\l_0$; moreover, $\LL
\bar\BB^{(0)}$ can be written as in \pref{h2}, with $Z_0^{(2)}= \tilde
Z_0^{(2)}=0$ (since no term proportional to $\psi^+_{\xx,\o} \psi^-_{\xx,-\o}$
can be present) and $Z_0^{(1)}$, $\tilde Z_0^{(1)}$ replaced by two new
constants $Z_0^{(3)}$, $\tilde Z_0^{(3)}$. Hence, we can analyze \pref{221} as
we did for \pref{2z}, but now we have only one r.c., to be called $\tilde\l_j$,
and three renormalization constants, $\tilde Z_j$, $Z_j^{(3)}$ and $\tilde
Z_j^{(3)}$, taking the place of $Z_j$, $Z_j^{(1)}$ and $\tilde Z_j^{(1)}$,
respectively. It follows that $\tilde\l_j \to \tilde\l_{-\io}$, as $j\to -\io$,
with $\tilde\l_{-\io}$ an analytic function of $\tilde\l_0$, such that
$\tilde\l_{-\io}= \tilde\l_0 + O(\tilde\l_0^2)$. On the other hand,
$\tilde\l_0$ is an analytic function of $\l_{\io}$ and $\tilde\l_0=
\l_{\io}+O(\l_\io^2)$, see \cite{[17]}; hence there exists an analytic function
$h(w)$, such that, if $\l_\io$ is small enough,
\be
\tilde\l_{-\io}=h(\l_{\io})\;.
\ee
Moreover, the flow equations of the new renormalization constants can be
written as in \pref{ffgz}, \pref{ffg}, \pref{ffga}, with different functions
$\b_t^{(j)}$ and $\bar\b_t^{(j)}$, $t=z, (\r,3), (J,3)$. However, if we put
\be c=v_s\;, \ee
the functions $\b_t^{(j)}$ are the same as before, as a consequence of the
definitions \pref{gjth} and \pref{gth}. It is then an immediate consequence of
\pref{ne1}, \pref{ne} and \pref{lau11} that
\bea\lb{lau12th}
&&|\log_\g(\tilde Z_{j-1}/\tilde Z_j) - \eta_z(\tilde\l_{-\io}/v_s)| \le
C_\th \l^2 \g^{\th j}\;,\nn\\
&&|\log_\g(Z^{(3)}_{j-1}/ Z^{(3)}_j) -
\eta_{\r,1}(\tilde\l_{-\io}/v_s)| \le C_\th \l^2 \g^{\th j}\;,\\
&&|\log_\g( \tilde Z^{(3)}_{j-1}/ \tilde Z^{(3)}_j) -
\eta_{J,1}(\tilde\l_{-\io}/v_s)| \le C_\th\l^2 \g^{\th j}\;,\nn
\eea
where $\eta_z(w)$, $\h_{\r,1}(w)$ and $\h_{J,1}(w)$ are {\it exactly} the same
functions appearing in \pref{lau12}. Hence, if we choose $\l_\io$, given $\l$,
so that
\be \tilde\l_{-\io} = \l_{-\io}\;,\ee
which is possible if $\l$ is small enough, the critical indices in the spin or
in the continuum model are the same.

We have now to show that the parameters $Z$, $Z^{(3)}$ and $\tilde Z^{(3)}$ of
the continuum model (with $c=v_s$) can be chosen so that \pref{h10} is true. To
begin with, we prove that they can fixed so that, for any $j\le 0$,
\bea\lb{ffgb}
&&|Z_j - \tilde Z_j|\le C_\th |\l| \g^{{\th\over 2} j}\;,\\
&&|Z^{(1)}_j - Z^{(3)}_j|\le C_\th |\l| \g^{{\th\over 2} j} \virg |\tilde Z^{(1)}_j -
\tilde Z^{(3)}_j|\le C_\th |\l| \g^{{\th\over 2} j}\;.\nn
\eea
Let us prove the first bound. By using \pref{lau12} and \pref{lau12th}, we see
that there exist $b_j(\l)$, $b$, $\tilde b_j(\l)$ and $\tilde b$, such that
\be
Z_j=b_j(\l) \g^{-j\h_z} \virg \tilde Z_j=Z \tilde b_j(\l) \g^{-j\h_z}\;,
\ee
with $|b_j(\l)-b| \le C_\th |\l| \g^{\th j}$ and $|\tilde b_j(\l)- \tilde b|
\le C_\th |\l| \g^{\th j}$. Hence, since $\th-\h_z\ge \th/2$, for $\l$ small
enough,
\be
|Z_j - \tilde Z_j| = Z_j \left| 1 -{Z \tilde b_j(\l)\over b_j(\l)} \right|\le
C_\th |\l| \g^{{\th\over 2} j}\;,
\ee
provided that we choose $Z=b/\tilde b$. In the same way we can choose the
values of $Z^{(3)}$ and $\tilde Z^{(3)}$.

Note that the values of $Z^{(3)}$ and $\tilde Z^{(3)}$ are expected to be
different, even if the asymptotic behavior, as $j\to -\io$, of $Z^{(3)}_j$ and
$\tilde Z^{(3)}_j$ is the same. This follows from the fact that the
``remainder'' $r_j$ in the representation \pref{ne1} of the propagator breaks
the symmetry \pref{symm}, which the relation $\h_z=\h_{\r,1}$ is based on. This
expectation is confirmed by an explicit first order calculation, see Appendix
\ref{app3}; we see that $Z^{(3)}=1-a \l+O(\l^2)$ and $\tilde Z^{(3)}=1+a
\l+O(\l^2)$, with
\be a = {1\over 2 \pi v_s}[\hat v(0)- \hat v(2 p_F)]\;.
\ee
Note that this expression is in agreement with the identity \pref{ggha}, since,
at first order $\l_{-\io}=\l_\io= 2\l [\hat v(0)- \hat v(2 p_F)]$.

\*

In order to complete the proof of \pref{h10}, we use the representation of $\la
S^{(3)}_\xx S^{(3)}_{\bf 0}\ra_T$, given in \cite{[13]}, eq. (1.13), that is
\be
\la S^{(3)}_\xx S^{(3)}_{\bf 0}\ra_T=\cos(2p_F
x)\O^a(\xx)+\O^b(\xx)+\O^c(\xx)\;,
\ee
where the first two terms represent the leading asymptotic behavior, while
$\O^c(\xx)$ is the remainder. In \cite{[13]} we proved that, if $\th<1$ and $n$
is a positive integer, then
\be\lb{bb1a} |\partial^n \O^a(\xx)|\le {C_n\over |\xx|^{2X_+ +n}}\virg
|\O^c(\xx)|\le{C_\th\over|\xx|^{2+\th}}\;,
\ee
where $X_+=K$ is the critical index \pref{l}. Moreover, by definition (see \S
5.9 of \cite{[13]}), $\O^b_\xx$ is a sum of multiscale graphs containing only
$\l$-vertices with scale $\le 0$ and in which the the propagators $g^{(h)}_{\o}$
and the renormalization constants $Z_h$, $Z_h^{(1)}$, $0\ge h\ge j$, are
replaced by $g^{(h)}_{th,\o}$ and $Z_h^{(th)}$, $Z^{(th,1)}_h$. It can be
written ( see (5.39) and (5.43) of \cite{[13]}), as
\be \O^b(\xx)=\sum_{h=-\io}^0 \sum_{\o=\pm} \left[ {Z^{(1)}_h\over
Z_h} \right]^2 [g^{(h)}_{th, \o}(\xx)g^{(h)}_{th,\o}(-\xx)+ G^{(h)}(\xx)]\;,
\ee
where $G^{(h)}(\xx)$ is a function satisfying, for any $N>0$, the bound
\be\lb{Gh} |G^{(h)}(\xx)|\le C_N {\g^{2h}\over 1+[\g^h|\xx|^N]}\;. \ee
The Fourier transform of $\O^c(\xx)$ is continuous; the same is true for
$\cos(2p_F x)$ $\O^a(\xx)$, around $\pp=0$, thanks to the bound (6.45) of
\cite{[13]} (where $\kk=\pp - 2\pp_F$ is bounded for $\pp$ small).

On the other hand we can write
\be G^{0,2}_{th,\r,\r}(\xx)=\sum_{h=-\io}^0 \sum_{\o=\pm} \left[
{Z^{(3)}_h\over \tilde Z_h} \right]^2 [g^{(h)}_{th,
\o}(\xx)g^{(h)}_{th,\o}(-\xx)+ \bar G^{(h)(\xx)}]+ G_1(\xx)\;,
\ee
where $\bar G^{(h)}(\xx)$ satisfies a bound similar to \pref{Gh}, as well as
$G_1(\xx)$, which is given by graphs with at least one propagator of scale $\ge
1$. Using \pref{ne1}, \pref{ne} and \pref{ffgb}, we get
\be \left| \int d\xx e^{i\pp\xx}[\O^b(\xx)-G^{0,2}_{th,\r,\r}(\xx)] \right|
\le \sum_{h=-\io}^0 \g^{(2+\th)h}\int d\xx {C_N\over 1+(\g^h|\xx|)}\le C_1\;,
\ee
which proves \pref{h10}.

It remains to prove the three equations \pref{h10a}; let us consider the first.
If $0<\k\le |\pp|,|\kk'|,|\kk'-\pp|\le 2\k$, in \S 2.4 of \cite{[16]} (see
(2.63) of \cite{[16]}) the following bound was proved,
\be\lb{bvert}
\left| \hat G^{2,1}_\r(\kk'+ \pp_F^\o, \kk'+\pp+\pp_F^\o) \right| \le {C\over
\k^{2-2\h}}\;,\ee
which is of course valid even for $G^{2,1}_{\r,th}(\kk',\kk'+\pp)$. Moreover,
if we choose the parameters of the continuum model as before, we can show, by
using again \pref{ne1}, \pref{ne} and \pref{ffgb}, that the difference
$R(\kk',\kk'+\pp)$ between $\hat G^{2,1}_\r(\kk'+ \pp_F^\o, \kk'+\pp+\pp_F^\o)$
and $G^{2,1}_{\r,th}(\kk',\kk'+\pp)$ is given by a summable sum of terms, each
bounded by the r.h.s. of \pref{bvert} times a factor $\g^{\th j}$. On the other
hand, if $h_\k \=\log_\g(\k)$ is the scale of the external fermion propagators,
each term of the expansion must have at least one propagator of scale $h_0\le
h_\k$; see (2.61), (2.62) of \cite{[16]} for a more detailed description of the
expansion. Hence, we can write, for $j\ge h_k$, $\g^{\th j} = \k^\th \g^{\th
(j- h_\k)}$ and we can absorb the factor $\g^{\th (j- h_\k)}$ in the bound,
thanks to the short memory property. It follows that
\be\lb{h10aa} \left| \hat G^{2,1}_\r(\kk'+ \pp_F^\o, \kk'+\pp+\pp_F^\o)
-G^{2,1}_{\r,th}(\kk',\kk'+\pp) \right|\le C_\th{\k^\th\over \k^{2-2\h}}\;,
\ee
from which the first of \pref{h10a} is obtained; the second and the third of
\pref{h10a} are proved by similar arguments.

\appendix

\section{Derivation of the Ward Identities \pref{tt}}\lb{app1}

Let us define $\psi^\pm_{\xx,\o} =\psi^{[l,N]\pm}_{\xx,\o}$, $\r_{\xx,\o} =
\psi^+_{\xx,\o} \ps^-_{\xx,\o}$ and let us consider the functional \pref{vv1a}.
By proceeding as in \S 2.2 of \cite{[13a]}, we can show that, by performing in
\pref{vv1a} the change of the variables $\psi^\pm_{\xx,\o}\to e^{\pm
i\a_{\xx,\o}}\psi^\pm_{\xx,\o}$ , the following identity is obtained:
\be\lb{gWI}
0={1\over Z(J)} \int P_Z(d\psi) \big[ -Z D_{\bar\o} \r_{\xx,\bar\o} +Z \d
T_{\xx,\bar\o}\big] e^{-\VV^{(N)}(\sqrt{Z}\psi) + \sum_{\o}\int\! d\xx
J_{\xx,\o} \r_{\xx,\o}}\;,
\ee
where $D_\o = \dpr_0 +i\o \dpr_1$, $Z(J)= \exp[\WW(J)]$ and
\be
\d T_{\xx,\o} = {1\over (L\b)^2} \sum_{\kk^+\not=\kk^-} e^{i(\kk^+-\kk^- )\xx}
C_\o(\kk^+,\kk^-) \hat\psi^{+}_{\kk^+,\o} \hat\psi^{-}_{\kk^-,\o}\;,
\ee
\be C_\o(\qq,\pp) = [\c_{l,N}^{-1}(\pp)-1] D_\o(\pp)
-[\c_{l,N}^{-1}(\qq)-1] D_\o(\qq)\;.
\ee

We now perform one functional derivative with respect to $J_{\yy,\o}$ in the
r.h.s. of \pref{gWI}, then we put $J=0$ and we take the Fourier transform. By
some trivial algebra, we get the two identities, valid for $\pp\not=0$ and for
any $\t$:
\bea\lb{34a}
&&\hspace{-.5cm} D_\o(\pp) G_{\o,\o}(\pp)- \t\ \hat v_0(\pp)
D_{-\o}(\pp) G_{-\o,\o}(\pp) =R_{N,1}(\pp)\;,\\
&&\hspace{-.5cm}  D_{-\o}(\pp) G_{-\o,\o}(\pp) - \t\  \hat v_0(\pp) D_{\o}(\pp)
G_{\o,\o}(\pp)=R_{N,2}(\pp)\;,\nn
\eea
where
\be
R_{N,1}(\pp)={\partial^2 \WW_A\over\partial \a_{\pp,\o}\partial J_{-\pp,\o}}
\Big|_{J=\a=0}\;, \quad R_{N,2}(\pp)={\partial^2 \WW_A\over\partial
\a_{\pp,-\o}\partial J_{-\pp,\o}}\Big|_{J=\a=0}
\ee
and
\be\lb{h11a}
e^{\WW_A (\a,\h,J)} = \int\! P_Z(d\psi) e^{-\VV^{(N)}(\sqrt{Z}\psi)+ \sum_{\o}
\int\! d\xx\ J_{\xx,\o} \r_{\xx,\o}} e^{\lft[A_0-\t
A_{-}\rgt]\lft(\a,\psi\rgt)}\;,
\ee
with
\bea
A_0 (\a,\ps) &=&\sum_{\o=\pm}\int\! {d\qq\;d\pp\over (2\p)^4}\
C_\o(\qq,\pp)\ha_{\qq-\pp,\o}\hp^+_{\qq,\o}\hp^-_{\pp,\o}\;,\\
A_-(\a,\ps) &=& \sum_{\o=\pm}\int\! {d\qq\;d\pp\over (2\p)^4}\
D_{-\o}(\pp-\qq)\hat v_0(\pp-\qq) \ha_{\qq-\pp,\o}\hp^+_{\qq,-\o}
\hp^-_{\pp,-\o}\;.
\eea
Note that the terms proportional to $\t$ in \pref{34a} are obtained by adding
and subtracting them to the identities one really gets; they are in some sense
two counterterms, introduced to erase the local marginal parts of the terms in
the effective potential proportional to $\a_{\xx,\o} \r_{\xx,\o}$, produced by
contracting the vertex $A_0$ with one or more $\l$ vertices. As shown in
\cite{[16],[17]}, the introduction of a non local interaction (still gauge
invariant) in the continuum model, makes it possible to calculate them
explicitly. Hence, the proof of \pref{tt} is equivalent to the proof that, if
$\t=\l_\io/4\pi c$ and $\pp\not=0$, then
\be
\lim_{-l,N\to\io} R_{N,1}(\pp)=-{1\over 4\pi c Z^2} D_{-\o}(\pp)\virg
\lim_{-l,N\to\io} R_{N,2}(\pp)=0\;.
\ee
This result is achieved by using the technique explained in \S 4 of
\cite{[14]}, that we shall now briefly explain.

The functional $\WW_A$ is analyzed, as always, by a multiscale integration and
a tree expansion; we get
\be\lb{vvv11}
R_{N,1}(\pp)= -{1\over Z^2}\int {d\kk\over (2\pi)^2}C_\o (\kk,\kk-\pp) \hat
g_{\o,th}^{[l, N]}(\kk) \hat g_{\o,th}^{[l, N]}(\kk-\pp)+ \bar R_{N}(\pp)\;,
\ee
where $\bar R_{N}(\pp)$ is given by the sum over all graphs with at least one
$\l$ vertex, while the first term in \pref{vvv11} is the $0$ order
contribution, coming from the contraction of the vertex $\d T_{\xx,\o}$ with
the vertex $\r_{\yy,\o}$. It is easy to show that, if $\pp\not=0$,
\be
\lim_{-l,N\to\io} \int {d\kk\over (2\pi)^2}C_\o(\kk,\kk-\pp) \hat
g_{\o,th}^{(l,N)}(\kk) \hat g_{\o,\th}^{(l,N)}(\kk-\pp)= {1\over 4\pi c}
D_{-\o}(\pp)\;.
\ee
Hence, to complete the proof, we have to show that, if $\pp\not=0$, $\bar
R_{N}(\pp)$ and $R_{N,2}(\pp)$ vanish in the removed cutoffs limit, thanks to
the choice of the counterterm $\t A_-$. This result is obtained by a slight
extension of the analysis given in \S 4 of \cite{[14]} for a similar problem;
we shall give some details, for people who have read that paper.

First of all, the sum over the graphs, such that one of the fermionic fields in
$A_0$ or $A_-$ is contracted at scale $l$, can be bounded by $C
\g^{l}|\pp|^{-1}$, hence it vanishes as $l\to-\io$, if $\pp$ is kept fixed at a
value different from $0$. Moreover, the sum over the other graphs, called
$\tilde R_{1,N}(\pp)$, can be written as
\be
\tilde R_{1,N}(\pp)=\sum_{k=0}^N \hat K_\D^{(1;0;1)(k)}+O(\g^{-\th N})\;,
\ee
where $\hat K_\D^{(1;2m;s)(k)}$ are the kernels of the monomials with one $\a$
field, $2m$ $\psi$ fields and $s$ $J$-fields in the effective potential, after
the integration of the scales $N,N-1,...k$, while the last contribution comes
from the trees with the root at a negative scale. The kernel $\hat
K_\D^{(1;2m;s)(k)}$ can be decomposed as in Fig. 4.1 of \cite{[14]} (with the
analogue of the terms $d$ and $e$ missing and a wiggling line in place of the
two fermion external lines). By proceeding as in the proof of (4.33)-(4.41) of
\cite{[14]}, we can see that
\be
|\hat K_\D^{(1;0;1)(k)}|\le C|\l_{\io}|\g^{-k} \g^{-\th(N-k)}\;.
\ee
It follows that $\bar R_N(\pp)=0$ vanishes in the removed cutoffs limit; the
same is true for $R_{2,N}(\pp)=0$, as we can prove in a similar way.

\section{Commutation rules and Ward Identities}\lb{app2}

Let us consider the model \pref{z1} and let us introduce the density and the
current operators (see \eg \cite{[8]}):
\bea
\r_x &=& S_x^3 + \frac12 = a^+_x a^-_x \virg x\in Z \;,\nn\\
J_x &=& S^1_x S^2_{x+1}-S^2_{x} S^1_{x+1}= \frac{1}{2i} [a^+_{x+1}a^-_{x}-
a^+_{x}a^-_{x+1}]\equiv v_F j_x\;.
\eea
As it is well known, the functions $G^{2,1}_\r(\xx,\yy,\zz)$ and
$G^{2,1}_j(\xx,\yy,\zz)$ can be written as
\bea
G^{2,1}_\r(\xx,\yy,\zz) &=& <T[\r_\xx a^-_\yy a^+_\zz]>_{L,\b}\;,\nn\\
G^{2,1}_j(\xx,\yy,\zz) &=& <T[j_\xx a^-_\yy a^+_\zz]>_{L,\b}\;,
\eea
where $<\cdot>_{L,\b}$ denotes the expectation in the Grand Canonical Ensemble,
$T$ is the time-ordered product and
\be
\r_\xx = e^{x_0 H} \r_x e^{-x_0 H} \virg a^\pm_\xx = e^{x_0 H} a^\pm_x e^{-x_0
H}\;.
\ee

The above definition of the current is justified by the (imaginary time)
conservation equation
\be\lb{eqm}
{\partial \r_\xx\over \partial x_0}= e^{H x_0} [H,\r_x] e^{-H x_0}
=-i\dpr^{(1)}_x J_{\xx} \= -i [J_{x,x_0}-J_{x-1,x_0}]\;,
\ee
where an important role plays the fact that
\be [H,\r_x]=[H_T,\r_x] \virg H_T = -\frac12 [a^+_{x}a^-_{x+1}+
a^+_{x+1}a^-_{x}]\;,\ee
a property which is not true for $J_x$.

By using \pref{eqm} and some trivial calculation, one gets the identity
\bea
&&{\dpr\over \dpr x_0} G^{2,1}_\r(\xx,\yy,\zz)  = -i v_F \dpr^{(1)}_x
G^{2,1}_j(\xx,\yy,\zz) +\nn\\
&&+\d(x_0-z_0)\d_{x,z} G^2(\yy-\xx) - \d(x_0-y_0)\d_{x,y} G^2(\xx-\zz)\;.
\eea
Let us now take the Fourier transform of the two sides of this equations. The
renormalization group analysis described in this paper implies that we can
safely take the limit $L,\b\to \io$ of $\hat G^{2,1}_\r(\kk,\kk+\pp)$, if $\pp$
and $\kk-\pp_F^\o$ are different from zero. Hence we get the identity
\pref{xa1a}, under the conditions on the momenta of Lemma \ref{lm1}, for any
value of $\k$.

In the same way we derive a WI for the density-density correlations. First we
observe that $G^{0,2}_{\r,\r}(\xx,\yy) = <T[\r_\xx \r_\yy]>_{L,\b}$; then, by
using \pref{eqm}, we get
\be
{\dpr\over \dpr x_0} G^{0,2}_{\r,\r}(\xx,\yy)  = -i v_F \dpr^{(1)}_x
G^{0,2}_{j,\r}(\xx,\yy) + \d(x_0-y_0) <[\r_{(x,x_0)}, \r_{(y,x_0)}]>_{L,\b}\;,
\ee
where $G^{0,2}_{j,\r}(\xx,\yy)$ is defined in a way similar to
$G^{0,2}_{\r,\r}(\xx,\yy)$, that is by using the definition in the last line of
\pref{corrXYZ}, with $\tilde J_\xx$ in place of $J_\xx$. Let us now take the
Fourier Transform; since $[\r_{(x,x_0)}, \r_{(y,x_0)}]=0$, we get, in the limit
$L,\b\to\io$, under the conditions on the momenta of Lemma \ref{lm1}, the
identity:
\be\lb{1000} -i p_0  G^{0,2}_{\r,\r}(\pp) -i(1-e^{-ip}) v_F G^{0,2}_{j,\r}(\pp)
=0\;,
\ee
which implies \pref{jjj1}.

{\bf Remark - } The WI \pref{xa1a} and \pref{1000} could also be obtained by
doing in \pref{1z} the change of variables $\psi^\pm_\xx\to e^{\pm
i\a_\xx}\psi^\pm_\xx$ and by proceeding as in App. \ref{app1} for \pref{vv1a}.
However, in this case the analysis of the corrections is much easier, since the
ultraviolet problem involves only the $k_0$ variable; it is indeed very easy to
prove that the corrections vanish in the $M\to\io$ limit.

\section{First order calculation of $Z^{(3)}$ and $\tilde Z^{(3)}$} \lb{app3}

$Z^{(3)}$ is defined so that $\lim_{h\to-\io} Z^{(3)}_h / Z^{(1)}_h =1$, see
\pref{ffgb}. On the other hand, at the first order, $Z^{(1)}_h = 1 + \a_h$,
where $\a_h$ is the sum of the values of the two Feynmann graphs of Fig.
\ref{fig1}, calculated at $\pp=0$ and $\tilde\kk = \pp_F^{\o}=(0,\o p_F)$ (the
result is independent of $\o$).
\insertplot{300}{70}{\ins{38pt}{48pt}{$\kk$}
\ins{35pt}{15pt}{$\kk-\pp$}
\ins{80pt}{25pt}{$\pp$}
\ins{15pt}{25pt}{$\pp$}
\ins{105pt}{50pt}{$\tilde\kk$}
\ins{120pt}{30pt}{$\tilde\kk-\pp$}

\ins{165pt}{25pt}{$\pp$}
\ins{200pt}{48pt}{$\kk$}
\ins{200pt}{15pt}{$\kk-\pp$}
\ins{235pt}{35pt}{$\kk-\tilde\kk$}
\ins{265pt}{55pt}{$\tilde\kk$}
\ins{265pt}{15pt}{$\tilde\kk-\pp$}
}{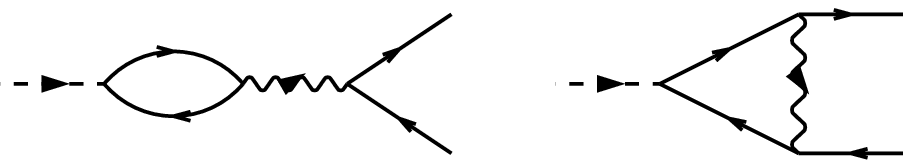}{\lb{fig1} The first order
contributions to the renormalization constants.}{0}

\0 By a simple calculation, we get, in the limit $M,L,\b\to \io$,
\bea\lb{alfa}
&&\a_h = -2\l \int {d\kk\over (2\p)^2} \hat g^{(\ge h)}(\kk)^2 [-\hat v(0) +
\hat v(k-\o p_F)] =\\
&&= -2\l \int_0^\p {dk\over (2\p)} \int_{-\io}^{+\io} {dk_0\over (2\p)} \hat
g^{(\ge h)}(\kk)^2 [-2\hat v(0) + \hat v(k-p_F) + \hat v(k+p_F)]\;,\nn
\eea
where $\hat g^{(\ge h)}(\kk) = \hat g^{(1)}(\kk) + \sum_{\o'} \sum_{j=h}^0 \hat
g_{\o'}^{(j)}(\kk-\pp_F^{\o'})$ is the propagator with infrared cutoff at scale
$h$, see \pref{gdef}. Note that, if $|k-\o' p_F|\ge \g^{h-1}$, $\hat g^{(\ge
h)}(\kk) = [-i k_0 + (v_s/v_F) (\cos p_F - \cos k)]^{-1}$ and that, if
$e_0\not=0$, $\int d k_0 [-i k_0 + e_0]^{-2} = 0$. It follows that, if
$\e=\g^h$,
\be
\a_h= - {\l [\hat v(0) - \hat v(2 p_F)]\over 2\p^2 v_s} \int_{-\e}^{\e} d t
\int_{-\sqrt{\e^2- t^2}}^{\sqrt{\e^2- t^2}} d k_0 {1\over (-i k_0 + t)^2} +
O(\e)\;,
\ee
so that
\be
\a_{-\io} = \lim_{h\to -\io} \a_h = - {\l [\hat v(0) - \hat v(2 p_F)]\over 2\p
v_s}\;.
\ee
A similar calculation can be done for $Z^{(3)}_h$; in fact, in this case, there
is no term corresponding to the second graph in Fig. \ref{fig1}, while the
contribution corresponding to the first one, with the external fermion
propagators of index $\o$, is given by
\be
\l \int {d\kk\over (2\p)^2} \hat g_{th,-\o}^{[h,N]}(\kk)^2\;,
\ee
with $g_{th,\o}^{[h,N]}(\kk)$ defined as in \pref{gth}. However, by the
symmetry \pref{symm}, the integral above vanishes for any $N$; hence, at the
first order, $Z^{(3)}_h=Z^{(3)}$, which implies that $Z^{(3)} = 1 + \a_{-\io} +
O(\l^2)$.

A similar procedure can be followed for the first order calculation of $\tilde
Z^{(3)}$. Let us consider first $\tilde Z^{(1)}_h$; since $v_F \hat j(\kk) =
\sin k\, a^+_\kk a^-_\kk$, we see immediately that $\tilde Z_h^{(1)} = 1 +
\lim_{h\to -\io} \o\b_{h,\o}$, where $\b_{h,\o}$ is obtained from \pref{alfa}
by inserting in the integrand a factor $\sin k/v_F$. It follows that
\be
\b_{h,\o} = -{2\l\over v_F} \int_0^\p {dk\over (2\p)} \sin k \int_{-\io}^{+\io}
{dk_0\over (2\p)} \hat g^{(\ge h)}(\kk)^2 [\hat v(k-\o p_F) - \hat v(k+ \o
p_F)]\;,
\ee
so that
\be
\lim_{h\to -\io} \o\b_{h,\o} = {\l [\hat v(0) - \hat v(2 p_F)]\over 2\p v_s} =
-\a_{-\io}\;.
\ee
On the other hand, we get as before that, at the first order, $\tilde Z_h^{(3)}
= \tilde Z^{(3)}$; hence $\tilde Z^{(3)} = 1 - \a_{-\io} + O(\l^2)$.

\section{Comparison with the Luttinger model}\lb{app4}

In the case of the Luttinger model, we can repeat the analysis leading to Lemma
\ref{lm1} and we can deduce two WI for the Luttinger model, which are similar
in the form to \pref{xa1}, \pref{xa3}. If we call $G^{2,1}_{L,\a,\o}$,
$\a=\r,j$, and $G^2_{L,\o}$ the correlation functions analogous to
$G^{2,1}_{th,\a,\o}$ and $G^2_{th,\o}$, we get the identities
\bea\lb{xaLutt}
&&-i p_0\ \hat G^{2,1}_{L,\r,\o}(\kk, \kk+\pp) + \o p\ \tilde v_J
\hat G^{2,1}_{L,j,\o}(\kk, \kk+\pp) =\nn\\
&&={Z^{(3)}\over (1-\t)Z} \left[ \hat G^2_{L,\o}(\kk) - \hat
G^2_{L,\o}(\kk+\pp) \right ] [1+O(\k^\th)]\;,\nn\\
&&-i p_0\ \hat G^{2,1}_{L,j,\o}(\kk, \kk+\pp) + \o p\ \tilde v_N
\hat G^{2,1}_{L,\r,\o}(\kk, \kk+\pp) =\\
&&={\tilde Z^{(3)}\over (1+\t)Z} \left[ \hat G^2_{L,\o}(\kk) - \hat
G^2_{L,\o}(\kk+\pp) \right ] [1+O(\k^\th)]\;,\nn
\eea
$\tilde v_J$ and $\tilde v_N$ being defined as in \pref{bert}. On the other
hand, exact WI for the Luttinger model can be obtained from the anomalous
commutation relations, see \eg \cite{[8]}. In our notation, we can write, if
$\s=\l_L/(2\p v_F)$ and $\l_L$ is the Luttinger coupling,
\bea\lb{h11Lutt}
&&-i p_0 \hat G^{2,1}_{L,\r;\o}(\kk,\kk+\pp)+ \o v_F p (1-\s)
\hat G^{2,1}_{L,j;\o}(\kk,\kk+\pp)]=\nn\\
&&\hspace{1cm} = \hat G^{2}_{th;\o}(\kk) - \hat G^{2}_{th;\o}(\kk+\pp)\;,\\
&&-i p_0 \hat G^{2,1}_{th,j;\o}(\kk, \kk+\pp) + \o v_F p (1+\s)
\hat G^{2,1}_{L,\r;\o}(\kk,\kk+\pp)] =\nn\\
&&\hspace{1cm} = \hat G^{2}_{th;\o}(\kk) - \hat G^{2}_{th;\o}(\kk+\pp)\;.\nn
\eea
By comparing \pref{xaLutt} with \pref{h11Lutt}, we get:
\be \lb{vva} \tilde v_J=v_s {Z^{(3)}\over \tilde Z^{(3)}}=v_F(1-\s)
\virg \tilde v_N=v_s {\tilde Z^{(3)}\over Z^{(3)}}=v_F(1+\s)\;, \ee
and
\be\lb{vvb} {Z^{(3)}\over(1-\t)Z}=1\virg
{\tilde Z^{(3)}\over (1+\t)Z}=1\;. \ee
The first identity in \pref{vvb} implies, as in the quantum spin chain case,
that $\k=K/ (\pi v_s)$. Moreover, the identities \pref{vva} imply that
\be
\lb{vvc}v_s=v_F\sqrt{(1-\s^2)}\;,
\ee
while \pref{vvb} and \pref{ggha} imply that
\be
{Z^{(3)}\over \tilde Z^{(3)}}={1-\t \over 1+\t}= K\;,
\ee
the relation between $K$ and $\t=\l_\io/(4\p v_s)$ being the same as in the
quantum spin model. On the other hand, \pref{vva} and \pref{vvc} imply also
that ${Z^{(3)}\over \tilde Z^{(3)}}= \sqrt{1-\s\over 1+\s}$; hence we have an
explicit expression of $K$ in terms of $\s$, that is:
\be\lb{exprK} K=\sqrt{1-\s\over 1+\s}\;. \ee

Note that \pref{vvc} and \pref{exprK} allow us to represent explicitly $v_s$
and $K$, which depend only on the large distance behavior of the model, in
terms of the ``bare'' quantities $\l_L$ and $v_F$. This result is strictly
related to the second identity in \pref{vvb}, which is missing in the spin
model, where it is replaced by the identity $\tilde v_J=v_F$, see \pref{hh}.
For the same reasons, the above equations imply also that, in the Luttinger
model, the following identities are true,
\be\lb{b111}
\tilde v_N=v_s K^{-1}\virg \tilde v_J=v_s K\;.
\ee
Note that these relations are also verified by the quantities $v_J$ and $v_N$,
introduced by Haldane in \cite{[6]}, but they are certainly {\it not true} in
the spin model model \pref{z}. In fact, in the $XYZ$ case one has, from the
second of \pref{hh}, that $\tilde v_J$ is $\l$-independent, while $v_s K$ is is
not, as it is evident from \pref{bb1} and \pref{bb2}. The relation \pref{b2} is
however valid also for the lattice model \pref{z}.

\end{document}